\documentclass{nature}
\usepackage[T1]{fontenc}
\usepackage[utf8]{inputenc}
\usepackage{graphicx}
\usepackage{pdflscape}
\usepackage{hyperref}
\usepackage{threeparttable}
\usepackage{xcolor}
\usepackage{ulem}

\usepackage{caption}
\usepackage[caption = false]{subfig}
\usepackage{academicons}
\usepackage{amsmath, amssymb}

\newcommand{\apj}{Astrophys. J.}
\newcommand{\apjl}{Astrophys. J.}
\newcommand{\apjs}{Astrophys. J.}
\newcommand{\aap}{Astron. Astrophys.}
\newcommand{\mnras}{Mon. Not. R. Astron. Soc.}
\newcommand{\nat}{Nature}
\newcommand{\araa}{Ann. Rev. Astron. Astrophys.}

\definecolor{orcidlogocol}{HTML}{A6CE39}

\def\be{\begin{eqnarray}}
\def\ee{\end{eqnarray}}

\makeatletter
\renewcommand{\baselinestretch}{1.2} 

\let\saved@includegraphics\includegraphics
\AtBeginDocument{\let\includegraphics\saved@includegraphics}
\renewenvironment*{figure}{\@float{figure}}{\end@float}
\makeatother

\makeatletter
\def\@fnsymbol#1{\ensuremath{\ifcase#1\or \dagger\or \ddagger\or
 \mathsection\or \mathparagraph\or \|\or **\or \dagger\dagger
 \or \ddagger\ddagger \else\@ctrerr\fi}}
\makeatother

\newcommand{\EXTTAB}[1] {Extended Data Table~\ref{#1}}
\newcommand{\EXTFIG}[1] {Extended Data Figure~\ref{#1}}
\newcommand{\SUPFIG}[1] {Supplementary Figure~\ref{#1}}

\usepackage[utf8]{inputenc}
\usepackage{hyperref}
\usepackage{graphicx}
\usepackage{longtable}
\usepackage{supertabular}
\usepackage{float} 

\usepackage{aas_macros}

\newcommand{\target}{4U 1543$-$47}
\newcommand{\arcdeg}{\mbox{$^{\circ}$}}
\newcommand{\arcmin}{\mbox{$^{\prime}$}}
\newcommand{\arcsec}{\mbox{$^{\prime\prime}$}}

\title{Jets from a stellar-mass black hole are as relativistic as those from supermassive black holes}

\author{
X. Zhang$^{1,6}$\href{https://orcid.org/0000-0002-8086-4049}{\includegraphics[scale=0.08]{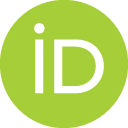}},  
W. Yu$^{1}$\href{https://orcid.org/0000-0002-3844-9677}{\includegraphics[scale=0.08]{ORCIDiD.png}}\thanks{E-mail: wenfei@shao.ac.cn}, 
F. Carotenuto$^{2}$,
R. Fender$^{2}$\thanks{E-mail: rob.fender@physics.ox.ac.uk},
S. Motta$^{3}$,
A. Bahramian$^{4}$,
J. C. A. Miller-Jones$^{4}$,
T. D. Russell$^{5}$, 
S.Corbel$^{7}$,
P. A. Woudt$^{8}$,
P. Atri$^{9}$, 
C. Knigge$^{10}$,
G. R. Sivakoff$^{11}$,
A. K. Hughes$^{11}$,
J. van den Eijnden$^{12}$,
J. Matthews$^{2}$,
M. C. Baglio$^{3}$
and P. Saikia$^{13}$}
\begin{document}
\maketitle
\begin{affiliations}

 \item Shanghai Astronomical Observatory, Chinese Academy of Sciences, Shanghai 200030, China
 
 \item Astrophysics, Department of Physics, University of Oxford, Keble Road, Oxford OX1 3RH, UK

 \item INAF-Osservatorio Astronomico di Brera, Via Bianchi 46, I-23807 Merate (LC), Italy

 \item International Centre for Radio Astronomy Research - Curtin University, GPO Box U1987, Perth, WA 6845, Australia

\item INAF, Istituto di Astrofisica Spaziale e Fisica Cosmica, Via U. La Malfa 153, I-90146 Palermo, Italy

 \item University of Chinese Academy of Sciences, Chinese Academy of Sciences, Beijing 100049, China 

 \item Université Paris-Saclay, Université Paris-Cité, CEA, CNRS, AIM, F-91191 Gif-sur-Yvette, France

 \item Department of Astronomy, University of Cape Town, Private Bag X3, Rondebosch 7701, South Africa

 \item ASTRON, Netherlands Institute for Radio Astronomy, Oude Hoogeveensedijk 4, NL-7991 PD Dwingeloo, the Netherlands

 \item School of Physics and Astronomy, University of Southampton, University Road, Southampton SO17 1BJ, UK

 \item Department of Physics, University of Alberta, CCIS 4-181, Edmonton, AB T6G 2E1, Canada

 \item Department of Physics, University of Warwick, Coventry CV4 7AL, UK

 \item Center for Astro, Particle and Planetary Physics, New York University Abu Dhabi, PO Box 129188, Abu Dhabi, UAE

 \end{affiliations}

\bigskip

\begin{abstract} 

Relativistic jets from supermassive black holes in active galactic nuclei are amongst the most powerful phenomena in the universe, acting to regulate the growth of massive galaxies. 
Similar jets from stellar-mass black holes offer a chance to study the same phenomena on accessible observation time scales. However, such comparative studies across black hole masses and time scales remain hampered by the long-standing perception that stellar-mass black hole jets are in a less relativistic regime. We used radio interferometry observations to monitor the Galactic black hole X-ray binary \target\ and discovered two distinct, relativistic ejections launched during a single outburst. Our measurements reveal a likely Lorentz factor of $\sim$ 8 and a minimum of 4.6 at launch with 95\% confidence, demonstrating that stellar-mass black holes in X-ray binaries can launch jets as relativistic as those seen in active galactic nuclei. 

\end{abstract}

\bigskip

\section*{Introduction}
The first superluminal jets discovered from the Galactic X-ray binary GRS~1915$+$105 some thirty years ago\cite{1994Natur.371...46M} proves the existence of microquasars \cite{1992Natur.358..215M} in our Universe. This opened up the window to study relativistic jets launched from nearby ($<10$ pc) accreting black holes in the Galaxy. Past observations have shown that the most relativistic jets measured in Galactic black hole X-ray binaries are those episodic jets launched when X-ray spectral transitions between the hard state (or intermediate state) and the soft state \cite{2004ARA&A..42..317F}, especially in black hole binary transients during the rising phase of an accretion outburst. The highest minimum Lorentz factor of relativistic jets constrained in these black hole X-ray binaries from those radio interferometric observations in the past few decades has been around $\sim$ 2.4 \cite{2004MNRAS.347L..52G,1995Natur.375..464H}. Since on average only 2-3 Galactic black hole X-ray binaries turn into bright outbursts yearly and some sources have turned into an outburst multiple times in the past\cite{2015ApJ...805...87Y}, the sample of stellar-mass black holes observed with episodic, relativistic jets is still limited to a few dozen sources to date.

Relativistic jets in Active Galactic Nuclei (AGN) are amongst the most powerful and extreme phenomena in the Universe that play a significant feedback role in regulating the evolution of massive galaxies \cite{2007ARA&A..45..117M}. These jets from supermassive black holes are much more relativistic (with Lorentz factors typically 10) \cite{2017ApJ...846...98J,2019ApJ...874...43L,2019ARA&A..57..467B} than those from their stellar-mass counterparts (with Lorentz factors assumed widely to be 2) \cite{2004ARA&A..42..317F}. The Lorentz factor of relativistic jets from AGN, based on a sample size more than an order of magnitude larger than those of the stellar-mass black holes, is found distributed in a broader range but more or less peaked slightly below $\sim$ 10 \cite{2017ApJ...846...98J,2019ApJ...874...43L,2019ARA&A..57..467B}. Observational studies of supermassive black holes tend to select face-on systems called blazars in a flux-limited manner, thus it has been widely believed that jets in stellar-mass black holes are not as relativistic, and they are laboratories of jet physics in a less relativistic regime\cite{2004ARA&A..42..317F}. But the actual Lorentz factor distribution of the jets in AGN is probably of a power-law form \cite{1997ApJ...476..572L}, suggesting that there are more samples towards lower Lorentz factors than what has been observed. On the other hand, jets from stellar-mass black holes in X-ray binaries offer a chance to study the same phenomenon but on timescales a million or more times shorter. Detection of similarly relativistic jets in stellar-mass black hole X-ray binaries and tracing them in full cycles of accretion events can therefore shed light on our understanding of relativistic ejections in the high-$\Gamma$ regime seen in AGN jets, as well as the entire process of building up the relativistic jet ejections that is otherwise inaccessible on current human-observable time scales.

\target~(IL Lup) is a black hole X-ray binary discovered in 1971 by the {\it Uhuru} satellite \cite{1972_ApJ}, with subsequent X-ray outbursts occurring nearly every decade \cite{1984_PASJ,1998ApJ...499..375O}. The system has a reported orbital inclination angle ${\theta}_{\rm orb}$ of 20.7$^{\circ}\pm$1.5$^{\circ}$ \cite{1998ApJ...499..375O,2003_IAUS_Orosz} with distance estimates of $d=$7.5$\pm$0.5~kpc from optical photometry \cite{2002_AAS_Orosz} and $d=$5.0$^{+2.0}_{-1.2}$ kpc from combining the parallax listed in Gaia Data Release 3 (Gaia DR3) and a Milky Way density prior for low-mass X-ray binaries \cite{2019MNRAS.489.3116A}. With a companion star of spectral type A2V and mass of 2.45$\pm$0.15 M$_{\odot}$, the system is thought to host a black hole with a mass ${M}_{\rm BH}$ of 9.4$\pm$1.0 M$_{\odot}$ \cite{2003_A&A_Ritter}, with the black hole spin $a_{*}$ estimated in the range of $0.67-0.98$ \cite{2023ApJ...946...19D,2020MNRAS.493.4409D,2006ApJ...636L.113S}. It went into an X-ray outburst in the summer of 2021. 
With regular radio interferometry observations, we discovered two sets of highly-relativistic, apparently superluminal jets in the black hole X-ray binary. We were able to trace the motion of the ejection blobs of these jets over about 15 months. The proper motions of the blobs are the highest ever detected in microquasars, implying a most likely Lorentz factor of 8 (with a 90\% confidence interval of 4.6-20.4) at launch, comparable to those seen in face-on AGN. 

\section*{Results}
\subsection{Radio and X-ray monitoring observations\\}

On June 11, 2021 (MJD 59,376), \target~was detected in a new X-ray outburst by the Monitor of All-sky X-ray Image (hereafter {\textit{MAXI}~\cite{2009PASJ}). On June 19 (MJD 59,384), we started weekly radio observations of the source of 15 minutes each in the 1.28 GHz band using the Meer Karoo Array Telescope (MeerKAT~\cite{2009IEEEP_Jonas}) under Large Survey Project (LSP) ThunderKAT \cite{fender2017thunderkat} (see Methods). These observations typically reached a sensitivity level of $\sim$~20 $\mu$Jy/beam. X-ray monitoring observations with {\it MAXI} and the Burst Alert Telescope (\textit{BAT}) on board the Neil Gehrels \textit{Swift} Observatory\cite{2004ApJ_Gehrels} (hereafter, \textit{Swift}) indicate that the black hole X-ray binary quickly transitioned from the X-ray hard state to the soft state on the same day as the outburst discovery. Quenching of the compact jet emission and the ejection of episodic transient jets have typically been associated with such X-ray spectral and timing state transitions in X-ray binaries\cite{2004ARA&A..42..317F}. Coordinated quasi-simultaneous observations by the X-ray Telescope (\textit{XRT}) on board \textit{Swift} and X-ray monitoring observations by \textit{MAXI}, \textit{Swift/BAT} and \textit{NICER} helped us discriminate X-ray states throughout the major outburst and subsequent mini-outbursts (see Figure~ \ref{fig:monitoring} and Methods). 

The first MeerKAT radio detection occurred on June 19, 2021 (MJD 59,384), spatially coincident with the known source position (Gaia DR2 optical position; see Methods). The radio emission remained consistent with this source ``core" position for the next three weeks. A new radio component $\sim$ 9$^{\prime\prime}$ away from the source was detected on September 5, 2021 (MJD 59,462), 84 days after the discovery of the X-ray outburst, indicating a transient jet had been launched at an earlier time. 

We continued our radio monitoring observations to track the downstream jet. The radio emission from the system is consistent with jet emission expected in the conventional picture \cite{2004MNRAS.355.1105F}. The downstream ejecta were detected in our MeerKAT observations of the source in the flux density range from 65$\pm$19 $\mu$Jy to 365$\pm$23 $\mu$Jy over a total of 24 epochs from September 5, 2021 (MJD 59,462) to January 7, 2023 (MJD 59,951), when we ceased the observations. All the detected ejecta were consistent with two ejections approaching the observer towards the South-East (SE) with consistent position angles within $\sim$1 $^{\circ}$ (see Methods).

\subsection{Astrometric measurements of discrete blobs\\}

Astrometric measurements up to November 8, 2021 (MJD 59,526) show that the first ejection (hereafter Ejection 1 or E1) underwent a clear deceleration when it reached an angular distance of 9--12$^{\prime\prime}$ away from the source, corresponding to a projected physical separation of $\sim$~0.2--0.3 pc at a source distance $d=5.0$\,kpc. According to the established picture of microquasar jets \cite{2004MNRAS.355.1105F}, E1 was likely launched after the rapid state transition during the early rise of the outburst, as there is no evidence of X-ray activity above its quiescent level up to more than two months prior to the discovery of the outburst (see Methods). E1 eventually reached an angular distance of $\sim$ 12 $^{\prime\prime}$ from the central binary, suggesting that the ejecta may have gone through a relatively low-density region before hitting a higher-density region $-$ at least $\sim$ 3 $^{\prime\prime}$ (or of 0.1 pc) thick along the direction of the proper motion (see Figure \ref{fig:propermotion}). We also identified a second ejection, hereafter Ejection 2 or E2, in observations from MJD 59,559 (see Methods). E2 followed the same path as E1; however, either due to E2 being more energetic or E1 having cleared a path, E2 propagated with a higher apparent speed to a much larger projected distance, reaching nearly 50$^{\prime\prime}$ ($\sim$ 1.2 pc) away from the central binary about 15 months later (see Figure \ref{fig:ejecta} and Methods). 

\subsection{Proper motions and constraints on Lorentz factors\\}

We model E1 as having a constant deceleration starting from a launch date that we conservatively set to June 11, 2021 (MJD 59,376), when \textit{MAXI} and \textit{Swift/BAT} observed that the source swiftly transitioned from the X-ray hard state to the soft state; E1 had an initial proper motion of 152.8$\pm$8.8 mas/day. For a source distance of 5 kpc, the Lorentz factor $\Gamma$ is constrained to $>$ 4.5$\pm$0.2 while the jet inclination angle ${\theta}_{\rm jet}$ should lie in the range 0$^{\circ}$ -- (25.6$^{\circ}\pm$ 1.4$^{\circ}$). We found the proper motion of E2 to be more extreme. By applying a constant-deceleration model to E2, we constrained its launch time as MJD 59,506.9$\pm$5.2 with an initial proper motion of 186.7$\pm$5.3 mas/day. The corresponding Lorentz factor is constrained to $>$ (5.5$\pm$ 0.2), while ${\theta}_{\rm jet}$ is constrained to lie in the range 0$^{\circ}$ -- (21.0$^{\circ}\pm$ 0.6$^{\circ}$). Both jet ejections are consistent with high Lorentz factors and a small jet inclination angle. 

The first detections of E1 and E2 in our MeerKAT monitoring campaign correspond to a projected angular separation about 9$^{\prime\prime}$ away from the central binary, likely suggesting there is a $\sim$~0.1 pc, or slightly larger inner cavity of interstellar media (ISM) surrounding the black hole X-ray binary. Before hitting the higher-density ISM further out, E2 was likely moving close to ballistic motion, since the previous E1 would have carved out a channel several months earlier. Thus we apply a ballistic model for E2 before its first detection on MJD 59,559. A constant-deceleration model of E2 is applied to ejecta beyond the cavity (see Figure \ref{fig:ejecta}). This provides a more conservative estimate of the launching date than our previous estimate using the constant deceleration model. The launch date of E2 is then obtained as MJD 59,502.7$\pm$5.0, with an initial proper motion of 166.8$\pm$2.7 mas/day. The launch date is well after the transition to the soft X-ray state, about 20 days after a few hard X-ray flares decayed, as seen with \textit{Swift/BAT}, suggesting an important role of the approaching disk flow in the ballistic jet launch (see Figure~\ref{fig:monitoring}). The corresponding Lorentz factor is then conservatively constrained to $>$ (4.9$\pm$0.1). This is the absolute minimum of the Lorentz factor at launch, dependent only on the assumed distance estimate of 5 kpc. Similarly, the ${\theta}_{\rm jet}$ is also constrained in the range 0$^{\circ}$ -- (23.5$^{\circ}\pm$ 0.4$^{\circ}$). The highest minimum $\Gamma$ at launch previously determined in all black hole X-ray binaries has been $\sim$ 2.4, seen in the stellar-mass black hole transients GX~339$-$4 and GRO J1655$-$40\cite{2004MNRAS.347L..52G,1995Natur.375..464H}. Our constraint on the minimum $\Gamma$ at launch in \target (see Methods) is a factor of two larger; the relativistic jets we detected have the most extreme minimum Lorentz factor observed to date in any microquasar. 
Furthermore, the proper motion is also significantly larger than the largest observed in microquasars in the past \cite{2021MNRAS.504..444C}.

Since the uncertainties in the estimates of the source distance and the jet inclination angle would affect the best constraint we can put on the Lorentz factor $\Gamma$, we made use of the prior distribution obtained from the Gaia DR3 parallax and assume a Milky Way density prior \cite{2019MNRAS.489.3116A} to perform Markov Chain Monte Carlo (MCMC) analysis to obtain the posterior distribution (see Methods). The inferred probability distribution of the Lorentz factor is shown in Figure \ref{fig:mcmc-E2}, which shows that the median is $7.8^{+5.0}_{-2.3}$, with a minimum of 4.6 at launch with 95\% confidence. This implies that the relativistic ejection E2 has a probability distribution of the Lorentz factor similar to the observed distribution of the Lorentz factors of the highly relativistic jets in more than a hundred radio-loud blazars \cite{2019ApJ...874...43L} and a few tens of gamma-ray blazars \cite{2017ApJ...846...98J,2022ApJS..260...12W}. 

The constraint on the \textit{minimum} Lorentz factor in \target~was derived from the astrometric measurements of the ejecta more than 50 days after launch, which when scaled by the black hole mass corresponds to an accretion timescale on the order of ${10}^{4-8}$ years for black holes with a mass ${\rm M}_{\rm BH} \sim {10}^{6-10}~{\rm M}_{\odot}$. If we were able to detect the ejecta soon enough and trace the ejecta at much earlier times, a higher Lorentz factor may have been detected, since the ISM in the vicinity of \target~, although could be thin, should still slow down the ejecta. The usual monthly to yearly monitoring time scales of very long baseline interferometry (VLBI) observations of AGN jets roughly correspond to accretion time scales of seconds in a microquasar. This implies that the relativistic jets in microquasars could be more extreme at early times than what we reported here. Future sensitive, high-cadence (hourly or less) VLBI monitoring observations may be able to trace jet motion in a more extreme regime with the help of new techniques such as dynamic phase centre tracking~\cite{2021MNRAS.505.3393W}. 

\section*{Discussion}

Since the first superluminal jets were discovered from GRS~1915$+$105 \cite{1994Natur.371...46M}, it has been accepted that jets from stellar-mass black holes are less relativistic than those from AGN \cite{2004ARA&A..42..317F}. 
However, our observations of \target~show jets from stellar-mass black holes can cover nearly the full range of the Lorentz factors observed in AGN jets, allowing us to probe similarly extreme regime of relativistic jet ejections in supermassive black holes as well. The small inferred inclination angle of the jets leads to a very high minimum Lorentz factor that would have been impossible to measure at larger inclination angles given our astrometric precision limits. Considering that only a few dozen Galactic stellar-mass black holes have been observed superluminal jets in the past few decades, this offers a natural explanation as to why this is the first measurement of a Lorentz factor comparable to those seen in AGN that occurred for this source. These jets were observed with a low inclination angle would likely be compatible with a scenario in which a substantial fraction of the microquasar jets are this relativistic. Furthermore, \target, as an outstanding example, showed two full cycles of accretion-ejection processes during a single X-ray outburst; consistent jet angles on different occasions indicate that those discrete jets may follow a direction fixed in time, for instance, the rotation axis of the black hole and/or the inner disk. Due to the mass-scale invariance of black hole accretion, the accretion and jet evolution time scales in black hole X-ray binaries are millions of times shorter than those in AGN. This allows us to observe full cycles of distinct accretion and jet ejections on human observing time scales to determine what properties persist and what properties vary, which is hardly achievable in observations of AGN. 

\bigskip
\bigskip
\bigskip

\begin{addendum}
\item This work made use of data from MeerKAT Large Survey Project (LSP) -- The HUNt for Dynamic and Explosive Radio transients with MeerKAT (ThunderKAT \cite{fender2017thunderkat}) (see \url{http://www.thunderkat.uct.ac.za}). The MeerKAT telescope is operated by SARAO, which is a facility of the National Research Foundation, an agency of the Department of Science and Technology. We thank RIKEN, JAXA and the \textit{MAXI} team for making the \textit{MAXI} data available. We acknowledge the Swift Guest Observing Facility for making BAT data products available and the UK Swift Science Data Centre for building most of the \textit{Swift/XRT} products in this research. For MeerKAT data reduction, we acknowledge the use of the ilifu cloud computing facility -- www.ilifu.ac.za, a partnership between the University of Cape Town, the University of the Western Cape, the University of Stellenbosch, Sol Plaatje University, the Cape Peninsula University of Technology and the South African Radio Astronomy Observatory. The ilifu facility is supported by contributions from the Inter-University Institute for Data Intensive Astronomy (IDIA - a partnership between the University of Cape Town, the University of Pretoria and the University of the Western Cape), the Computational Biology division at UCT and the Data Intensive Research Initiative of South Africa (DIRISA). ATCA is part of the Australia Telescope National Facility (\url{https://ror.org/05qajvd42}) which is funded by the Australian Government for operation as a National Facility managed by CSIRO. We acknowledge the Gomeroi people as the Traditional Owners of the ATCA observatory site. We thank Jamie Stevens and ATCA staff for scheduling the observations. W.Y. and X.Z. acknowledge support from the Natural Science Foundation of China (No. U1838203 and 12373050). F.C. acknowledges support from the Royal Society through the Newton International Fellowship programme (NIF/R1/211296). G.R.S. is supported by NSERC Discovery Grant RGPIN-2021-04001.

\item[Author contributions] 
X.Z. and W.Y. analyzed MeerKAT data and performed model fits;  W.Y. and R.F. proposed the interpretation; F.C. and R.F. independently cross-checked the MeerKAT results; X.Z., S.M. and W.Y. performed X-ray data analysis; A.B., J.M.J. and P.A. investigated source distances and optical measurements; F.C. and S.M. performed the MCMC analysis and tests; T.D.R. carried out and analyzed the ATCA data; C.K. performed cross-checks of model fits; R.F. and P.W. initiated the MeerKAT Large Survey Project (LSP) ThunderKAT campaign; X.Z. and  W.Y. contributed to the primary draft; all authors contributed to the writing and discussion of the paper. 


\item[Correspondence] Correspondence and requests for materials
should be addressed to W. Yu (wenfei@shao.ac.cn) and R. Fender (rob.fender@physics.ox.ac.uk).

\item[Data availability]
MeerKAT data are directly available from the MeerKAT archive. 
\textit{MAXI} and \textit{Swift/BAT} X-ray monitoring data are also in the public domain. This research has made use of \textit{Swift/XRT} and \textit{NICER} data, which are publicly available and can be obtained through the High Energy Astrophysics Science Archive Research Center (HEASARC) website at \url{https://heasarc.gsfc.nasa.gov/W3Browse/}.

\item[Code availability]
Much analysis for this paper has been undertaken with publically available codes and the details required to reproduce the analysis are contained within the manuscript.

\end{addendum}

\clearpage

\begin{figure}[H]
 \centering
 \includegraphics[angle=0,width=6.0in]{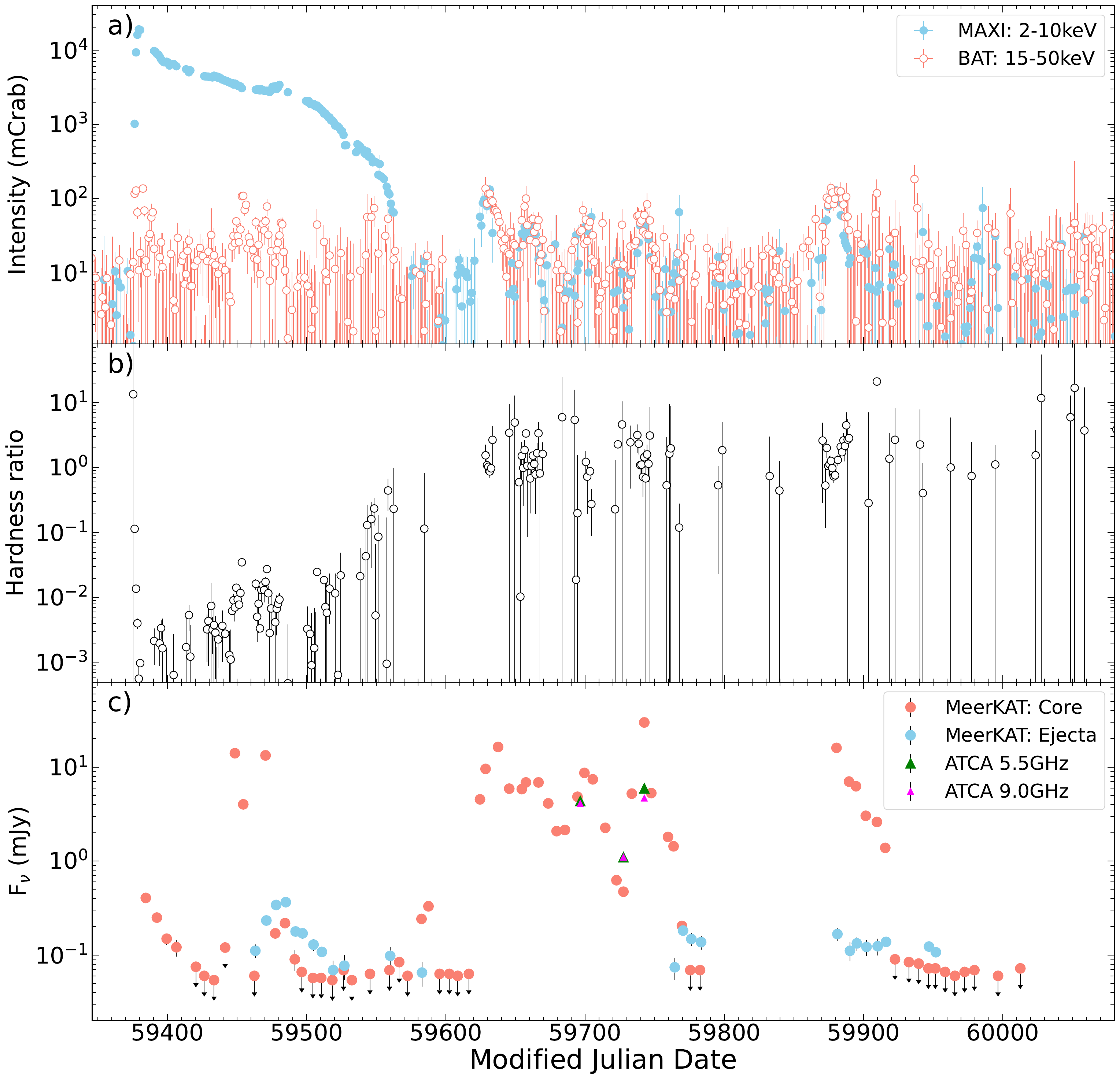}
 \caption{
 \footnotesize \textbf{Long-term evolution of the source in the X-ray and radio bands as seen with MAXI, Swift/BAT and MeerKAT.} a) The X-ray daily-averaged light curves from \textit{MAXI} (blue) and \textit{Swift/BAT} (red). b) The hardness ratio defined as the 15--50 keV intensity with \textit{BAT} in Crab units over the 2--10 keV intensity with \textit{MAXI} in Crab units. c) MeerKAT and ATCA radio flux density measurements of the core (in red) and the ejecta (in blue) during the 2021--2023 outburst of \target. The measurements of the core emission will be presented in a separate paper.}
 \label{fig:monitoring}
\end{figure}

\begin{figure}[H]
 \centering
 \includegraphics[angle=0,height=14cm]{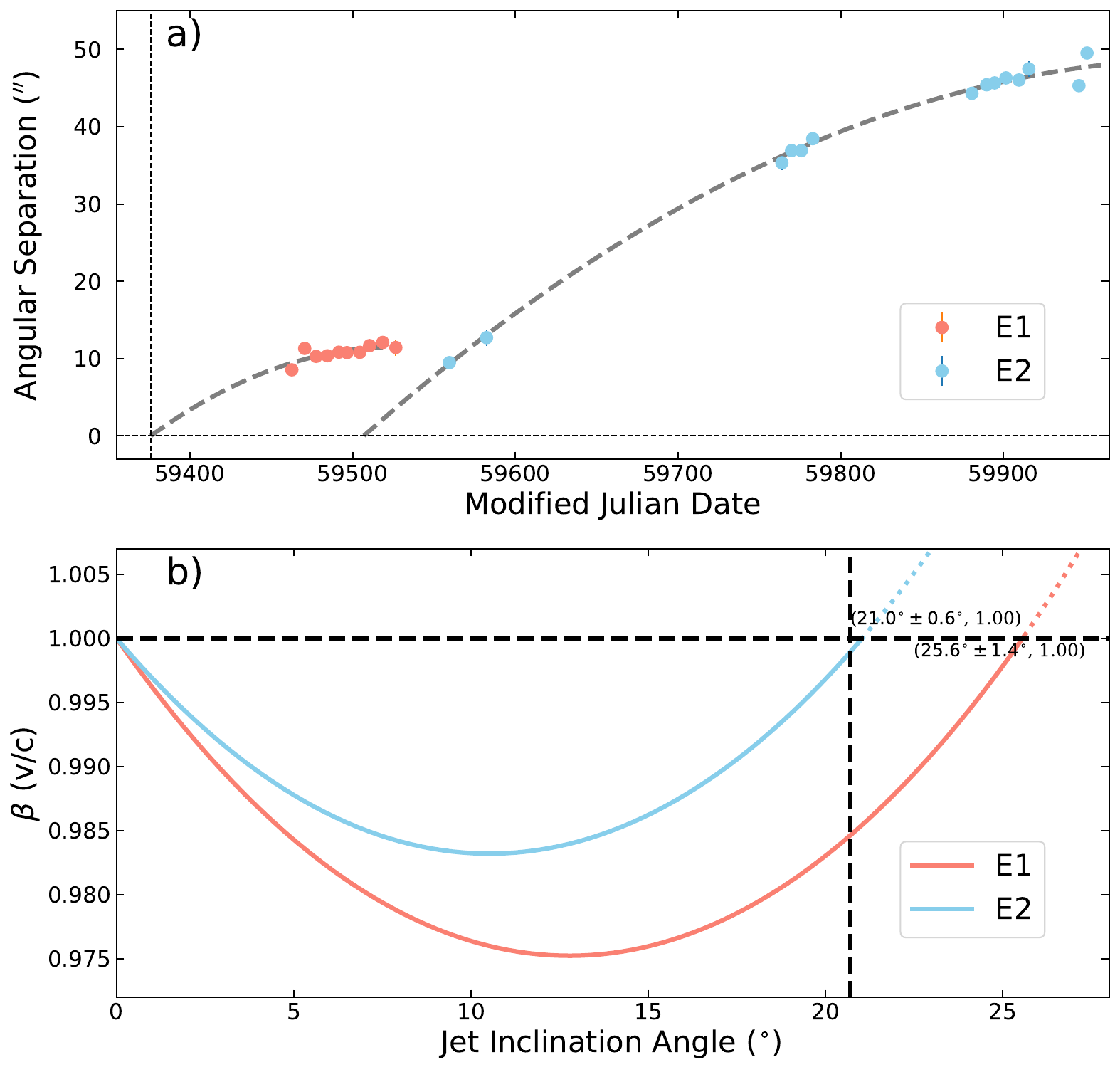}
 \caption{
 \footnotesize \textbf{ The evolution of the angular separation between the ejection blobs and the core and the inferred constraints on $\beta$ and the jet inclination angle.} a) The separation between the blob and the core with best-fit models (dashed lines) over-plotted. The ejecta detected up to MJD 59,526 correspond to E1 (in red), while those detected afterwards correspond to E2 (in blue). The vertical dotted line marks the start time of the 2021--2023 outburst. The dashed line represents a constant-deceleration model of the proper motion of the ejecta corresponding to E1 and a ballistic proper motion plus a constant-deceleration model of that of the discrete blobs corresponding to E2. b) Allowed ranges of $\beta$ and the jet inclination angles by E2 (blue solid line) and E1 (red solid line). The vertical dashed line marks the likely orbital inclination angle of the binary from optical observations ${\theta}_{orb}$ of 20.7$^{\circ}\pm$1.5$^{\circ}$. The proper motion of the discrete blobs corresponding to E2 constrains the jet inclination angle to the range 0$^{\circ}$-- (21.0$^{\circ}\pm$0.6$^{\circ}$) and the minimum of $\beta$ to above $\sim$~0.983~c ($\Gamma$ = 5.5$\pm$0.2); the proper motion of the discrete blobs corresponding to E1 constrains the jet inclination angle to the range 0$^{\circ}$-- (25.6$^{\circ}\pm$1.4$^{\circ}$) and the minimum of $\beta$ to above $\sim$~0.975~c ($\Gamma$ = 4.5$\pm$0.2). Both indicate the jet ejections are nearly face-on.}
 \label{fig:propermotion}
\end{figure}

\begin{figure}[H]
 \centering
 \includegraphics[width=12cm]{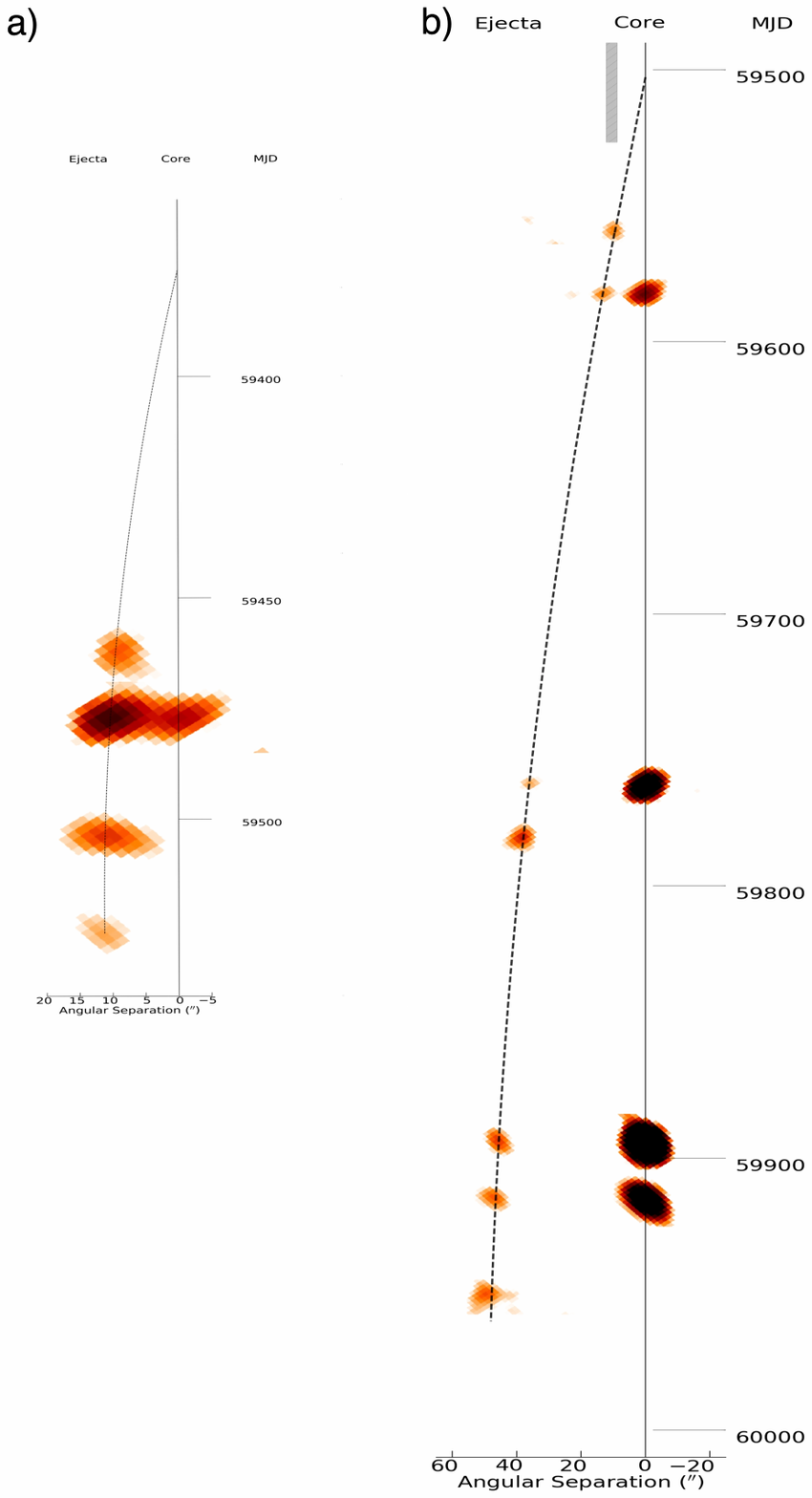}
 \caption{
 \footnotesize \textbf{ The schematic picture of the proper motion of E1 (a) and E2 (b) in relative to the core, shown with selected radio images from the MeerKAT observations.} The images have been vertically shifted by an amount proportional to the corresponding time spans between the observations. The shaded gray region at the top of b) represents the angular range of E1 relative to the core, consistent with the ejecta angular range shown in a). The dashed lines show the best-fit models for E1 and E2, respectively. The model of the proper motion of E2 is composed of an initial ballistic motion up to the first detection on MJD 59,559 with subsequent constant deceleration. }
 \label{fig:ejecta}
\end{figure}

\begin{figure}[htbp]
 \centering
  \includegraphics[width=0.65\textwidth,angle=0]{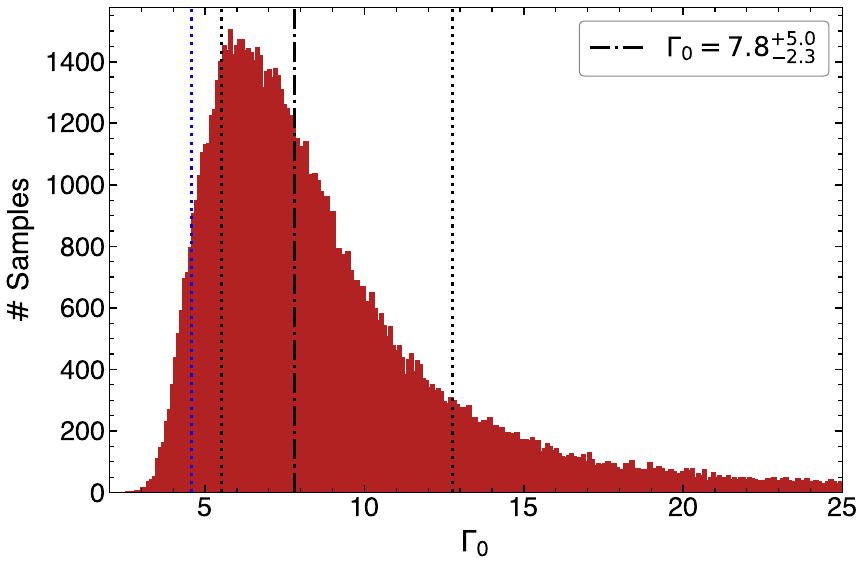}
 \caption{
 \footnotesize \textbf{Posterior distributions for the Lorentz factors derived for E2.} The results are obtained by randomly sampling from the distributions of source distance, proper motion and inclination angle (see Methods). The posterior distribution of the initial Lorentz factor $\Gamma_{0}$ for E2 is obtained with the launch date free to vary. The legend shows the median value of $\Gamma_{\rm 0} = {7.8}^{+5.0}_{-2.3}$ (black dot-dashed vertical line) along with these 1$\sigma$ uncertainties (vertical black dotted lines), reported as the difference between the median and the 16th percentile of the posterior (lower error bar), and the difference between the 84th percentile and the median (upper error bar). The vertical blue dotted line marks the value of $\Gamma_{\rm 0}$ above which we find the 95\% of the samples. The median and the range of the probability distribution of the Lorentz factor match the range of the Lorentz factors of highly relativistic jets observed in radio-loud \cite{2019ApJ...874...43L} and gamma-ray \cite{2022ApJS..260...12W} blazars.}
 \label{fig:mcmc-E2}
\end{figure}


\newpage 

\begin{methods}

\section*{X-ray Observations and Data Reduction}
\subsection{MAXI and Swift/BAT monitoring observations\\}
The daily and orbital \textit{MAXI} and \textit{Swift/BAT} X-ray monitoring light curves are publicly available. We obtained the \textit{MAXI} light curves in the 2--4~keV, 4--10~keV, and 10--20~keV bands\footnote{\url{http://maxi.riken.jp/top/slist.html}}, respectively. We also obtained the \textit{Swift/BAT} 15--50 keV light curves\footnote{\url{https://swift.gsfc.nasa.gov/results/transients/BlackHoles.html}}. We use these X-ray light curves to identify source states, and soft and hard X-ray flares, as well as to plot the figures in the main text and this section. The decrease in MAXI flux from MJD 59,454.0 to 59,463.0, and from 59,381.0 to 59,390.0, were not used due to \textit{MAXI} detector structure and shadows by the \textit{MAXI} detector structure, respectively.
 
\subsection{Swift/XRT and NICER pointed observations\\}
There are 30 successful \textit{XRT} observations (observation 00014374020 was not successful) of \target~during the 2021--2023 outburst. We extracted the energy spectrum per observation using the UK \textit{Swift} Science Data Centre pipeline, which yielded good X-ray spectra \cite{2007A&A_Evans,2009MNRAS_Evans} except for the observation 00089352003. For this observation, we had to manually extract source and background spectra, as well as the  {\it mkf} and {\it arf} files using the {\it xrtmkarf} task. We then fitted each spectrum in the 1--10 keV band with XSPEC \cite{1996ASPC_Arnaud} with either a power-law model or disk blackbody model (\textit{powerlaw} and \textit{diskbb} in XSPEC), or a combination of the two components, multiplied by an interstellar absorption using \textit{tbfeo}. We froze the Galactic neutral hydrogen absorption column density to $\rm N_H=0.338\times{10^{22}\,{\rm  cm}^{-2}}$ $-$ the Galactic value in the direction of our source \cite{2016A&A_HI4PI} in all the spectral fits. The unabsorbed fluxes in the 1--10 keV band were obtained with the task {\it cflux}. These results were used to identify the X-ray state and will be reported in greater detail in a separate paper reporting on the radio vs.\ X-ray correlation in the source.  

\target~was monitored intensely by \textit{NICER} during its 2021--2023 outburst since June 12, 2021, only one day after the \textit{MAXI} discovery of the X-ray outburst. We reduced the \textit{NICER} data using the pipeline tool {\it nicerl2} with standard filtering and the Ftools tool \textit{XSELECT} to extract spectra. The background was estimated using the tool nibackgen3C50, which implements the 3C50 background model \cite{Remillard_2022}. The Focal Plane Module (FPM) No.~14 and No.\ 34 were removed because of detector noise. The response matrix files (RMF) and ancillary files (ARF) were generated with the Ftools tool {\it nicerrmf} and {\it nicerarf}.  Similar to the XRT data reduction, we fitted the spectra in 1--10 keV range with XSPEC using the same models. The Galactic neutral hydrogen absorption column density was fixed to $\rm N_H=0.338\times{10^{22}\,{\rm  cm}^{-2}}$ throughout the fit as what we did for XRT data. The unabsorbed fluxes in the 1--10 keV band were also obtained with {\it cflux}. These measurements were used to identify X-ray state. Specifically, we found the source marginally entered and stayed in the X-ray intermediate state (which lies between hard and soft states) on MJD 59,453.

\section*{Radio Observations and Data Reduction}
\label{sec:Radio_obs} 
\subsection{MeerKAT observations and data reduction\\}
\label{sec:meerkat_data}
MeerKAT is a precursor array telescope of the Square Kilometre Array (SKA). The telescope in the current phase consists of 64 antennas, each with a diameter of 13.5 meters, and has a maximum baseline of 8 km. 
As part of ThunderKAT, which targets X-ray binaries, Cataclysmic Variables, Supernovae, and Gamma-Ray Bursts, we monitored the stellar-mass black hole transient 4U 1543$-$47 under the X-ray binary monitoring program with MeerKAT on $\sim$ weekly cadence starting on June 19, 2021, about 8 days after the discovery of the 2021 X-ray outburst, and ending on March 9, 2023, when the source had already returned to quiescence. During the MeerKAT campaign, data were taken at a central frequency of 1.28 GHz with 32k channels in total reaching a bandwidth of 856 MHz, with integration times set to 8 seconds. We observed J1939$-$6342 for 5 minutes at the beginning of each run to calibrate the absolute flux and bandpass. J1501$-$3918 was used as the phase calibrator and was observed for 2 minutes before and after typical on-source exposure time of $\sim$~15 minutes on the target field. Standard data reduction procedures\footnote{\url{https://casaguides.nrao.edu/index.php?title=Karl_G._Jansky_VLA_Tutorials}} (i.e., flagging, calibration, applying solutions to data, splitting target data) were done through Common Astronomy Software Application package (CASA, version 5.6.2, hereafter CASA \cite{2007ASPC_McMullin}). Then we flagged additional calibrated target data in CASA to exclude bad data that still remained. Finally, we produced images in total intensity (Stokes I) using {\it tclean} in CASA after performing phase self-calibration of the data. We used a Briggs weighting scheme with a robust parameter of $0.0$ to achieve a compromised solution between minimizing the side-lobe effects across the field and maximizing sensitivity. A multi-scale multi-frequency synthesis algorithm was carried out in the image deconvolution while multi-frequency synthesis with {\it nterms}=2 and a central frequency of 1.28 GHz were set. Flux densities and positions for the ejections of \target~were measured through the CASA task {\it imfit}. Errors on flux densities were measured by taking the root-mean-square (RMS) noise of a close-by source-free region in each observation. The source detections and corresponding flux densities of the source were also cross-checked by both the dedicated Oxkat semi-automated pipeline \cite{2020ascl.soft_oxkat} and the Science Data Processor (SDP) pipeline images, all of which show consistency with our measurements as introduced above. The detailed MeerKAT measurements of the core jet emission and the investigation of the core jet emission physics will be presented in a separated paper. 

\subsection{ATCA observations and data reduction\\}
\target\ was observed by the Australia Telescope Compact Array (ATCA) on three dates during its 2021--2023 outburst, under project code CX501 (PI: T.D. Russell). Observations were performed on April 27, 2022 17:37:00 -- 22:03:40 UT (MJD~59,696.83$\pm$0.09), May 28, 2022 11:24:50 -- 13:33:30 UT (MJD~59,727.52$\pm$0.04), and June 12, 2022 09:16:10 -- 18:03:30 UT (MJD~59,742.57$\pm$0.18). During these observations, ATCA was in its more extended 6D, 1.5B and 6B configurations\footnote{\url{https://www.narrabri.atnf.csiro.au/operations/array_configurations/configurations.html}}, respectively. Data were recorded simultaneously at central frequencies of 5.5 GHz and 9.0 GHz, with 2 GHz of bandwidth at each central frequency. Data were flagged, calibrated, and imaged following standard procedures within \textsc{casa}\footnote{\url{https://casaguides.nrao.edu/index.php/ATCA_Tutorials}}(version 5.1.2). J1939$-$6342 was used for bandpass and flux calibration and the nearby (3.1$^\circ$) B1600$-$489 was used for complex gain calibration. Imaging followed the same procedure as outlined for the MeerKAT data 
, using a Briggs robust \cite{1995PhDT.......238B} parameter of 0. The flux density and position of the target was measured by fitting for a point source in the image plane with the \textsc{casa} task \textit{imfit}. Errors in the flux density are taken as the RMS noise of a close-by source-free region added in quadrature with conservative systematic uncertainties of 4\% at 5.5 and 9 GHz \cite{2010MNRAS.402.2403M,2016ApJ...821...61P}. These ATCA observations correspond to radio flux measurements of the core jet emission, which will be studied in detail in a separated paper. 

\section*{Radio Astrometry of the Ejection Blobs}
\label{sec:Pos_Ast} 
\noindent
We first detected a discrete blob that we associated with E1 at an angular separation of $\sim$~9$^{\prime\prime}$ from the core emission on September 5, 2021, with the 1.28~GHz image, and then clearly traced subsequent ejection blobs for E1 with monitoring observations to an angular separation of $\sim$~12$^{\prime\prime}$ from \target~over 60 days. The ejection displayed evidence of significant deceleration in the end. The later angular separation translates to a projected separation of $\sim$~0.3 pc at the source distance of 5~kpc. Then a month later, we discovered additional discrete blobs at a position closer to the core and that showed apparently higher proper motion than those of E1. This indicates an additional ejection (E2) happened. The ejection had remained detected until January 7, 2023, reaching an angular distance $\sim$~50$^{\prime\prime}$ away from \target, corresponding to $\sim$~1.2 pc at the source distance of 5~kpc. Examples of radio detections of the ejection blobs corresponding to either E1 or E2 and the core are shown in~\EXTFIG{fig:4panel_sky_im} and \SUPFIG{fig:contour_img}. The measured radio flux densities are listed in~\EXTTAB{tab:meerkat_observations}.  

To obtain accurate angular separations from the core for all radio emitting blobs, we made use of their measured radio positions and the Gaia DR2 optical position of \target~ (ICRF: [J2000] = 15h47m08.27667s, $-47$\arcdeg40\arcmin10.285\arcsec) as the position where the blobs originated. Absolute positions of the blobs in the radio band were measured by CASA {\it imfit} task. We evaluated the systematical offset of each position of the discrete blobs in the framework of Gaia DR2~\cite{2018A&A_GaiaDR2}, which provides the reference astrometry.  The offset was obtained by subtracting the radio position of each epoch from the Gaia DR2 optical position
(The International Celestial Reference Frame (ICRF) of (R.A., Dec.)~[J2000] = (15h47m18.8419s, $-47$\arcdeg37\arcmin40.306\arcsec)) of a bright check source (hereafter BCS), which is $\sim$~3$^{\prime}$ away from \target~ with a stable radio flux density of $\sim$~1 mJy. 

Instead of recording the statistical errors on R.A.\ and Dec.\ of the discrete blobs reported from CASA {\it imfit} task, which are thought to underestimate the statistical errors, we estimated the statistical errors of R.A.\ and Dec.\ of the discrete blobs by calculating the corresponding size of synthesized beam in the direction of R.A.\ and Dec.\ divided by twice the signal-to-noise ratio in each observation. 
Similarly, we obtained the statistical errors on R.A.\ and Dec.\ for each observation for the BCS. Then we added in quadrature the statistical errors on R.A.\ and Dec.\ of both sources to get the positional errors for all the discrete blobs. All of the positional measurements are shown in~\EXTTAB{tab:meerkat_observations}. Figure~\ref{fig:propermotion} shows the angular separation vs.\ time of all the discrete blobs we detected. 

\section*{Position Angle of the Ejecta}

Based on the errors on the R.A.\ and Dec.\ estimated above, we can also calculate the position angles for all the radio-emitting discrete blobs of the ejections, which are shown in \EXTFIG{fig:Position_Angle}. The fitted position angle of the blobs is 130.1$\pm$0.3$^{\circ}$ East of North. Following the classification of ejection blobs as belonging to two ejections E1 and E2 (see below), we obtained the position angles 128.9$\pm$0.9$^{\circ}$ and 130.3$\pm$0.3$^{\circ}$ East of North, for E1 and E2, respectively, suggesting that the two ejections were launched with consistent position angles. If E1 and E2 have identical paths, E1 would have swept the way for E2, allowing E2 to go further and faster at the same angular separation from its origin; it is also possible that E2 might have hit and passed through E1. Our measurements of the position angles of the ejections also indicate that episodic jets from the stellar-mass black hole had a surprisingly stable inclination angle during the outburst -- a specific jet inclination angle is preferred by the two ejections from \target. 

\section*{Distance Estimation}
\label{sec:dis_estimate} 

Previous optical observations have resulted in a best estimate of the distance of 7.5$\pm$0.5 kpc~\cite{2002_AAS_Orosz} to the black hole X-ray binary through modeling optical light curves and spectra of the system, which improved from a previous estimate of 9.1$\pm$1.1 kpc~\cite{1998ApJ...499..375O}. An independent estimate of the distance of the black hole X-ray binary can be derived from the parallax with a low significance of around 3.5~$\sigma$ (0.191$\pm$0.055 mas) measured with Gaia, as listed in DR3. Assuming Bailer-Jones' prior that relies on the Galactic stellar distribution, which BH Low Mass X-ray Binaries do not follow, a distance of 5$\pm$2 kpc is derived~\cite{2024arXiv240111931F}. While combining with the assumption of the prior that is derived based on the distribution of BH Low Mass X-ray Binaries~\cite{2019MNRAS.489.3116A}, a distance of 5$^{+2.0}_{-1.2}$ kpc (where a zero point correction has also been applied) is estimated (\EXTFIG{fig:distance}). The later distance is consistent with the previous best-estimate, and can be used to infer conservatively how relativistic the jet ejections would be.

\section*{Proper Motion of the Ejecta and the Modelings}

\subsection{Physical Considerations\\}
Proper motions of the ejecta can tell us about important properties of jets and source environment. Modeling of the proper motion will tell us about how relativistic the jet is, as well as the jet's inclination angle and launch time. In episodic jets of microquasars, each discrete blob corresponds to a forward-moving plasma clump that is launched from the source. A single ejection can interact with the ISM on its way forward, allowing us to detect them multiple times in both time and space sequences. It is worth noting that the following physical considerations are necessary to model the proper motion of the ejecta and helpful for the understanding.  

Firstly, an ejection blob will move further and further from the source position before it terminates.  We shall call it the "Going Forward" rule. This means that the radio position of the same ejection at a later time is more distant from the source than its previous position. This helped us identify the occurrence of a new ejection E2 and distinguish E2 from E1 during the outburst of \target. 

Secondly, any ejection blob can only move at most with a ballistic motion or decelerate due to its interactions with the ISM. We shall call this the "No Acceleration" rule. No acceleration of any episodic jet ejecta has been observed before in microquasars. In the angular separation vs.\ time plot of the ejection blobs, acceleration or deceleration will be shown as a steepening or a flattening of the slope. This is the second reason that E2 must have occurred. 

Thirdly, we should consider a modeling of the proper motions of the blobs with the least number of ejections possible. We shall call this the "Least Ejection" rule. This helps us get clues of jet physics with the most simplest interpretation but with the strongest constraints from data. We started from a single ejection model to fit the data and found it is not possible to model the angular separation vs.\ time relation. It helps us to model the ejecta in the simplest way to reveal the primary physics from the ejecta's proper motion. Notice that one can model every detection of ejecta as a single isolated ejection event, but clearly it will not be physical.  

\subsection{Classification of Ejections\\}
With the above considerations in mind, we now perform model fits to the angular separation vs.\ time relation shown in Figure~\ref{fig:propermotion}. At first, we try to model the motion of the discrete blobs with a single ejection. Both "Going Forward" rule and "No Acceleration" rule provide strong evidence against the single ejection scenario: (1) The blob detected on MJD 59,559, marked as Blob A, is at an angular distance 9.48 $\pm$ 0.82 $^{\prime\prime}$ from our source, significantly lower by 2.4 $\sigma$ than the averaged angular separations in the range 11.44 -- 12.09 $^{\prime\prime}$ the ejection reached more than two months ago during MJD 59,510--59,526. In addition, alternative model fits suggest Blob A significant deviate ($3.7\sigma$ or more) from the trend of the previous blob movement (see below); (2) the blobs during MJD 59,559--59,782 show proper motion with a higher apparent speed than that formed by the discrete blobs detected before MJD 59,526 in the angular separation vs.\ time plot, indicating the later discrete blobs, attributing to a single ejection, had a higher proper motion. Thus those blobs we detected should come from more than one ejection. 

We investigate the detected discrete blobs to identify those associated with E1. By default, all the discrete blobs in the group before MJD 59,526 are considered as members of the same ejection (E1). The second discrete blob detected on MJD 59,470 seems to deviate from the adjacent detections by showing a larger angular separation. Our investigation shows that the second discrete blob and the core emission are not fully distinguished in the image, bringing complexity in the determination of source components and positions (\EXTFIG{fig:4panel_sky_im}). The position and flux density of the second blob were obtained by fixing a Gaussian at the Gaia position. Therefore, though the second discrete blob shows a deviation, it still likely belongs to E1. 

Since Blob A detected on MJD 59,559 is found at a smaller angular separation than the previous blobs, it should correspond to a new ejection from \target. Furthermore, the Blob A, B, and others detected in the period MJD 59,559--59,782 establish a steeper slope in the angular separation vs.\ time plot. This demonstrates that these blobs should correspond to an additional ejection, namely E2, instead of E1. We noticed that there is a statistical possibility that Blob B belongs to E1, but this would require: (1) E1 remained at a slow proper motion for nearly two months and (2) E2 did not hit the relatively slowly moving E1 from behind so that E2 could maintain its motion and (3) E2 could not be detected when Blob B (the E1 blob) was detected. Therefore that the Blob B belongs to E2 is more reasonable. In summary, by considering the rules for ejection classification and various investigations of model fits, we associate all the blobs detected before MJD 59,526 as due to E1 and those detected since MJD 59,559 as due to E2. Detailed model fits and additional MCMC tests suggest that the angular separation vs.\ time relation of all the discrete blobs is very well explained with only two ejections $-$ E1 and E2 from \target~(see below).    

\subsection{Constraints on the Relativistic Ejections\\}
To estimate the intrinsic speed of the bulk motion of E1, we conservatively fixed the launched date to $t_{\rm ej}$ = MJD 59,376 when the conservative fast hard-to-soft state transition occurred at the beginning of the 2021 outburst. Then applying a constant deceleration model in the form of 

\begin{equation}
D = \mu(t-t_{\rm ej}) + \frac{1}{2}\dot{\mu}(t-t_{\rm ej})^{2}
\label{eq:equation_1}
\end{equation}

\noindent where $D$ is the measured angular separation, $\mu$ is the initial proper motion in mas/day, $\dot{\mu}$ is the deceleration in mas/day$^{2}$, and $t_{\rm ej}$ is the ejection date. In addition, we put constraints to make sure the modeled proper motion can never become negative. For E1, we obtain an initial proper motion of $\mu=$152.8$\pm$8.8 mas/day with the angular separations of the blobs ranging from $\sim$~0.2 pc to $\sim$~0.3 pc with a reduced $\chi^{2} \sim$~2.4 (dof$=$8). Applying the same simple model for E2, we obtain an initial proper motion of 186.7$\pm$5.3 mas/day with a reduced $\chi^{2} \sim$~2.4 (dof$=$11) while tracing its motion from $\sim$~0.3 pc to $\sim$~1.2 pc at a source distance of 5 kpc over a course of over one year, as shown in Figure~\ref{fig:propermotion}.

Below we solve the apparent velocity corresponding to E1 and E2. For a proper motion $\mu_{\rm app}$, the apparent velocity is expressed as a
fraction of the speed of light $\beta_{\rm app} = v_{\rm app}/c$, for
the source at a distance $d$ is,

\begin{equation}
\beta_{\rm app} \sim \left(\frac{D}{\rm kpc}\right) \left(\frac{\mu_{\rm app}}{\rm 173~mas/d}\right)
\label{eq:equation_2}
\end{equation}


Therefore it is striking that E2 moved initially with an apparent speed of $v_{\rm app}$ of $\sim$~5.4~$c$ when taking a distance to \target~as 5.0 kpc, or more strikingly, of $\sim$~8.1~$c$ when taking a distance of 7.5 kpc. 

We can also put constraints on the minimum of $\beta$. If a jet is propagating at an intrinsic speed $\beta$ at an inclination angle of $\theta$ for a source at a distance of $D$, then its proper motion is,

\begin{equation}
\mu_{\rm app} = \frac{\beta \sin\theta}{1 - \beta \cos{\theta}} \frac{c}{D},
\label{eq:equation_3}
\end{equation}

Therefore the dependence of the intrinsic speed of jet ejecta vs.\ jet inclination angle can be obtained by rearranging the equation above, 

\begin{equation}
\beta = \frac{{D}{\mu_{\rm app}}/{c}}{\sin\theta + \cos{\theta}({D}{\mu_{\rm app}}/{c})}
\label{eq:equation_4}
\end{equation}

Thus $\beta$ has a minimum value $\beta_{\rm min}$ when $\tan\theta$ = ${c}/{D}{\mu_{\rm app}}$. 

We can further put constraints on the minimum of $\Gamma$ and maximum jet inclination angle physically allowed by the measured proper motion. Given the measured proper motion of E2 and assuming a source distance of 5 kpc, the corresponding initial Lorentz factor is $\Gamma_{\rm min}$ = 5.5$\pm$0.2 ($\sim$ 0.983~$c$) when $\theta$ $\sim 10.5^{\circ}$. While the maximum allowed inclination angle $\theta_{max}$ is $\sim$ 21.0$^{\circ}\pm$ 0.6$^{\circ}$, otherwise $\beta$ will be $>$~1, which would be unphysical. Applying a constant deceleration model introduced above to those E2 blobs beyond its first detection, then extrapolating the slope of the first detection back to when the angular separation is zero, provides a more conservative estimate of the jet launching date and initial proper motion. The resultant launching date is thus estimated to MJD 59,502.7$\pm$5.0 and initial proper motion is 166.8$\pm$2.7 mas/day. This leads to a minimum initial Lorentz factor of 4.9$\pm$0.1, with ${\theta}_{\rm jet}$ only allowed in the range 0$^{\circ}$-- (23.5$^{\circ}\pm$ 0.4$^{\circ}$). Performing the same calculations to E1, we found an apparent speed of $\sim$~4.4~$c$, $\Gamma_{\rm min}$ = 4.5$\pm$0.2 ($\sim$~0.975~$c$) for arbitrary jet inclination angle in the allowed varying range of 0$^{\circ}$-- (25.6$^{\circ}\pm$ 1.4$^{\circ}$). These is by far the most relativistic lower limit to date for jets from Galactic stellar-mass black hole X-ray binaries.

\subsection{Distribution of the Lorentz factor with MCMC \\}

We have further investigated the jet motion for the ejections E1 and E2 with a simple constant deceleration model, where the parameters are the initial proper motion $\mu$ in mas/day and the deceleration ($\dot{\mu}$) in (mas/day)$^{2}$, as well as the ejection date~$t_{\rm ej}$. We used a Bayesian approach with MCMC (with the emcee python library), adopting a flat prior on $t_{\rm ej}$, $\mu$ and $\dot{\mu}$. For E2, we left all the parameters free to vary, while for E1 we fixed the ejection date to MJD 59,376. In the MCMC, we used 100 walkers that sampled the parameter space for a total of $10^5$ steps. The best fits are shown in \EXTFIG{fig:angsep_mcmc}, while the posterior corner plots for $\mu$, $\dot{\mu}$ and $t_{\rm ej}$ are shown in \EXTFIG{fig:mcmc-corner}. 

Then, using the posterior distributions on the proper motion (see the corner plots in \EXTFIG{fig:mcmc-corner}), we obtained posterior distributions on the initial Lorentz factor $\Gamma_0$ of the two ejections. This was done by randomly sampling 100000 times from the derived probability distribution of the distance, proper motion and inclination angle of the jets, and then computing the Lorentz factor from Equation~\ref{eq:equation_4}. For the source distance, we used the prior distribution obtained from the Gaia DR3 parallax and assuming a Milky Way density prior for Low Mass X-ray Binaries (\EXTFIG{fig:distance}).
For the inclination angle $\theta$, we assumed a flat prior in $\cos{\theta}$, in which $\theta$ varies between 1$^{\circ}$ and the maximum angle allowed from the combination of the sampled values of distance and proper motion. The probability distributions of $\Gamma_0$ are shown in \EXTFIG{fig:mcmc-gamma}.
Considering the median of the probability distributions, the $\Gamma_0$ obtained for the ejections E2 and E1 are 7.8$^{+5.0}_{-2.3}$ and 6.7$^{+4.4}_{-2.0}$, respectively. 
In particular, for E1 we find that 95\% of the samples are above $\Gamma_0 > 3.8$, while the same percentile for E2 yields $\Gamma_0 > 4.6$.

\bigskip

\end{methods}

\newpage

\section*{Extended Data}

\renewcommand{\baselinestretch}{1.0}
\selectfont

\noindent

\bigskip\noindent \EXTFIG{fig:4panel_sky_im}: {\bf Four sample images of the source and the ejecta as seen with MeerKAT.} The source position (yellow) and the ejecta (red for E1 and blue for E2) are marked with a cross. The image of the source and the ejecta obtained in the observation on MJD 59,470 is shown in a). The position of the ejecta, corresponding to the second detection of E1, can not be measured accurately due to confusion with the source in this observation.

\bigskip\noindent \EXTFIG{fig:Position_Angle}: {\bf The position angles of the ejections E1 and E2.} The position angle of each discrete emitting blob of both E1 and E2 are shown with red and blue filled circles, respectively.

\bigskip\noindent \EXTFIG{fig:distance}: {\bf Probability distribution for the distance of \target~ used in the MCMC analysis.} This is obtained from the Gaia DR3 parallax measurement while assuming the prior that derived based on BH Low Mass X-ray Binaries~\cite{2019MNRAS.489.3116A}.

\bigskip\noindent \EXTFIG{fig:classification}: {\bf The measurements of the angular separations of all the discrete blobs and the classification of E1 (red) and E2 (blue).} Blob A marks the first discrete blob of E2, showing a steeper slope with Blob B and later blobs than those blobs detected at earlier times, indicating the occurrence of E2 following E1.

\bigskip\noindent \EXTFIG{fig:angsep_mcmc}: {\bf Angular separation in arcsec between the ejecta and the position of 4U~1543--47 and the MCMC best-fit constant deceleration models for E1 and E2.} The best fits obtained with the constant deceleration models, with a fixed ejection date for E1, and a free ejection date for E2.

\bigskip\noindent \EXTFIG{fig:mcmc-corner}: {\bf Corner plots showing the constraints on the decelerated motion for the ejections.} The panels on the diagonal show histograms of the one dimensional posterior distributions for the model parameters, including the jet proper motion $\mu$ in mas/day, its time derivative $\dot{\mu}$ in mas/day$^2$ and the ejection date $t_{\rm ej}$, reported as $\sim$~MJD 59,506.

\bigskip\noindent \EXTFIG{fig:mcmc-gamma}: {\bf Posterior distributions for the Lorentz factors derived for E1.}

\bigskip\noindent \EXTTAB{tab:meerkat_observations}: {\bf Angular separations of all discrete blobs measured with MeerKAT Observations.}

\newpage
\renewcommand{\thefigure}{\arabic{figure} } 
\renewcommand{\thetable}{\arabic{table} } 
\renewcommand{\figurename}{\text{Extended Data $|$ Figure}}
\renewcommand{\tablename}{\text{Extended Data $|$ Table}}

\bigskip
\begin{figure*}
\centering
\includegraphics[width=0.70\textwidth,angle=0]{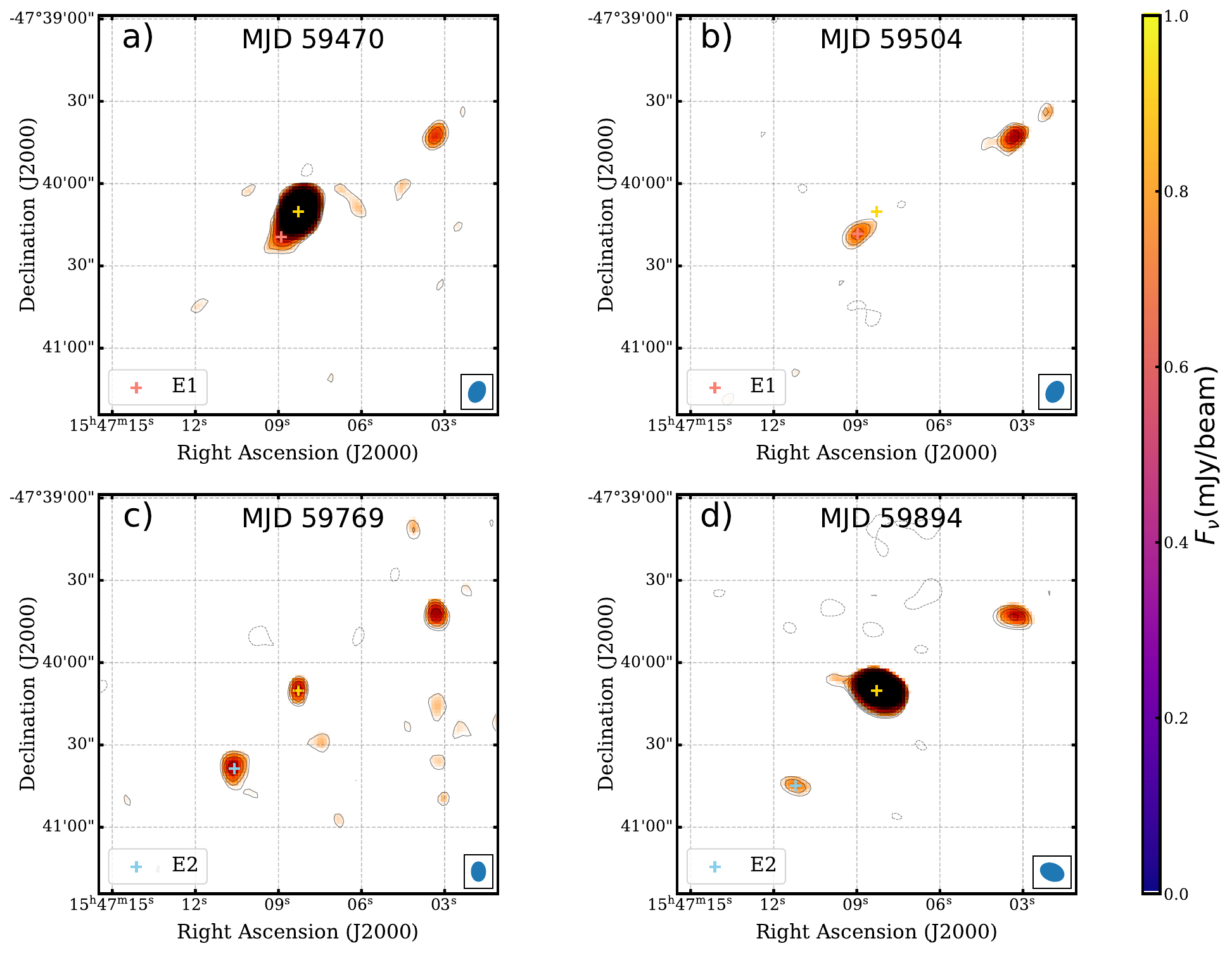}

\caption{
\footnotesize \textbf{Four sample images of the source and the discrete blobs as seen with MeerKAT.} The source position (yellow) and the discrete blobs (red for E1 and blue for E2) are marked with crosses. The image of the source and the discrete blob obtained in the observation on MJD 59,470 is shown in a). The position of the discrete blob can not be measured accurately due to confusion of the two components, causing obvious deviation of the estimated position from the trend for E1. b) to d) show clearly the two components, i.e., the core emission from the source at the center and the discrete blobs at the bottom left of the images. Contours correspond to -1 , 1, $\sqrt{2}$, 2, 3, 4, and 5 times the 3$\times$rms noise, respectively. }
\label{fig:4panel_sky_im}
\end{figure*}
\clearpage

\newpage

\bigskip

\begin{figure*}
\centering
\includegraphics[width=0.95\textwidth]{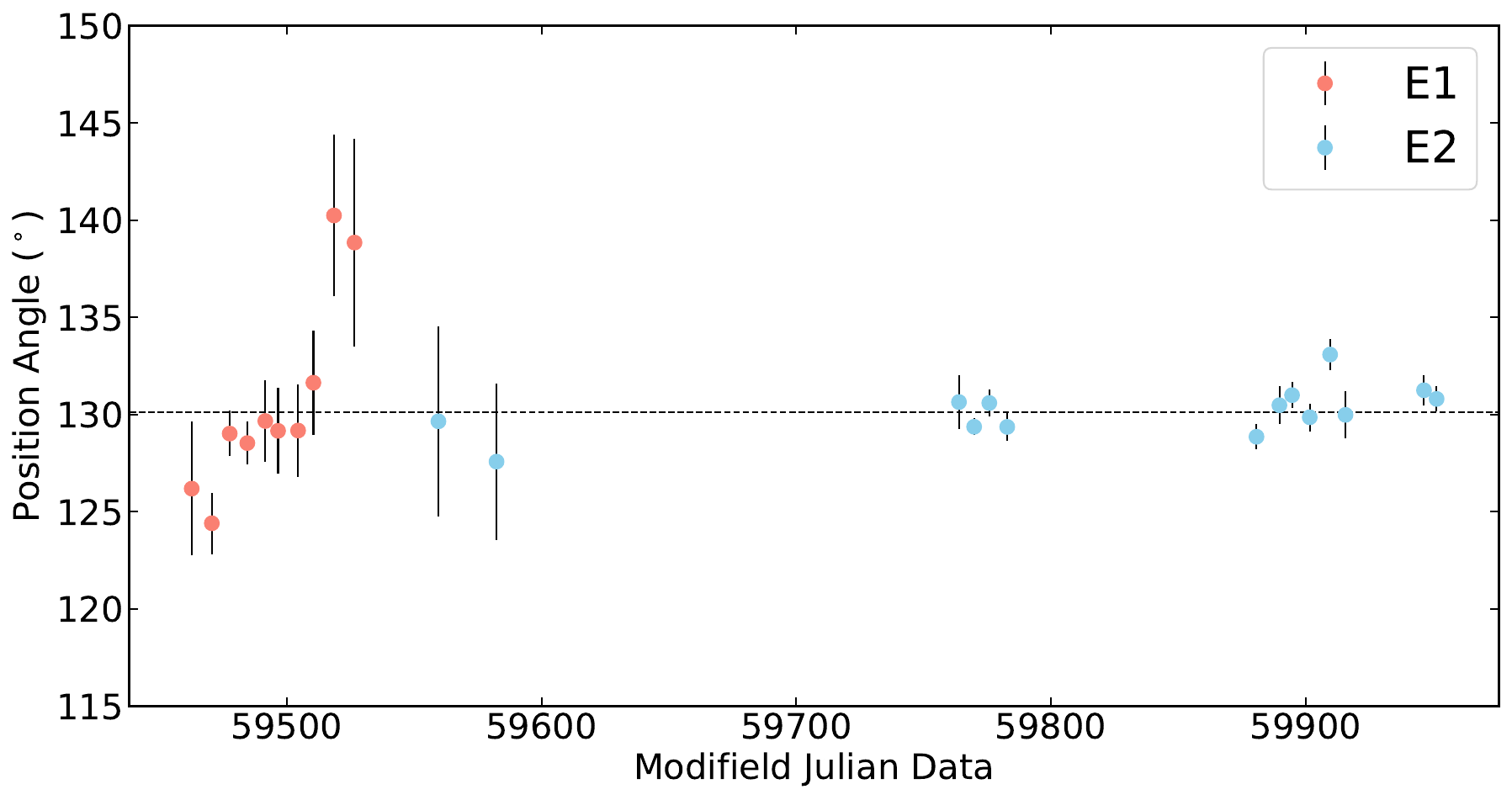}

\caption{
\footnotesize \textbf{The position angles of the ejections E1 and E2.} The position angle of each discrete emitting blob of both E1 and E2 are shown with red and blue filled circles, respectively. E1 and E2 can be fitted with a position angle of 128.9$\pm$0.9$^{\circ}$ and 130.3$\pm$0.3$^{\circ}$ East of North respectively. Jointly fit E1 and E2 results in a position angle of 130.1$\pm$0.3$^{\circ}$ East of North.}
\label{fig:Position_Angle}
\end{figure*}
\clearpage

\newpage
\bigskip
\begin{figure*}
 \centering
 \includegraphics[width=0.7\textwidth]{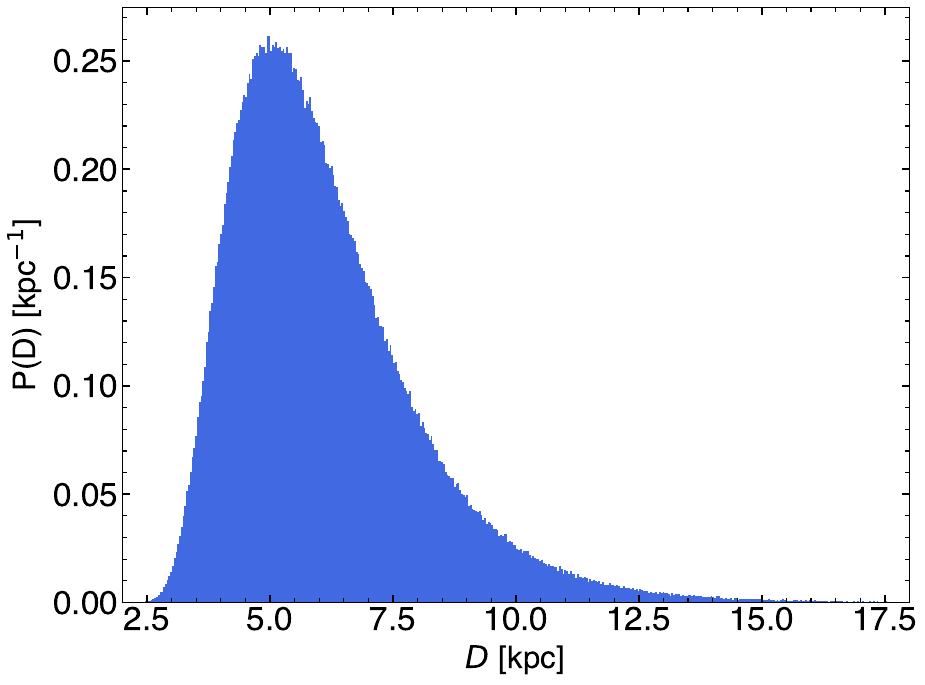}
 \caption{
 \footnotesize \textbf{Probability distribution for the distance of \target.} This is obtained from the Gaia DR3 parallax measurement while assuming the prior that derived based on BH Low Mass X-ray Binaries~\cite{2019MNRAS.489.3116A}.}
\label{fig:distance}
\end{figure*}
\clearpage

\newpage

\begin{figure*}
 \centering
 \includegraphics[width=\textwidth]{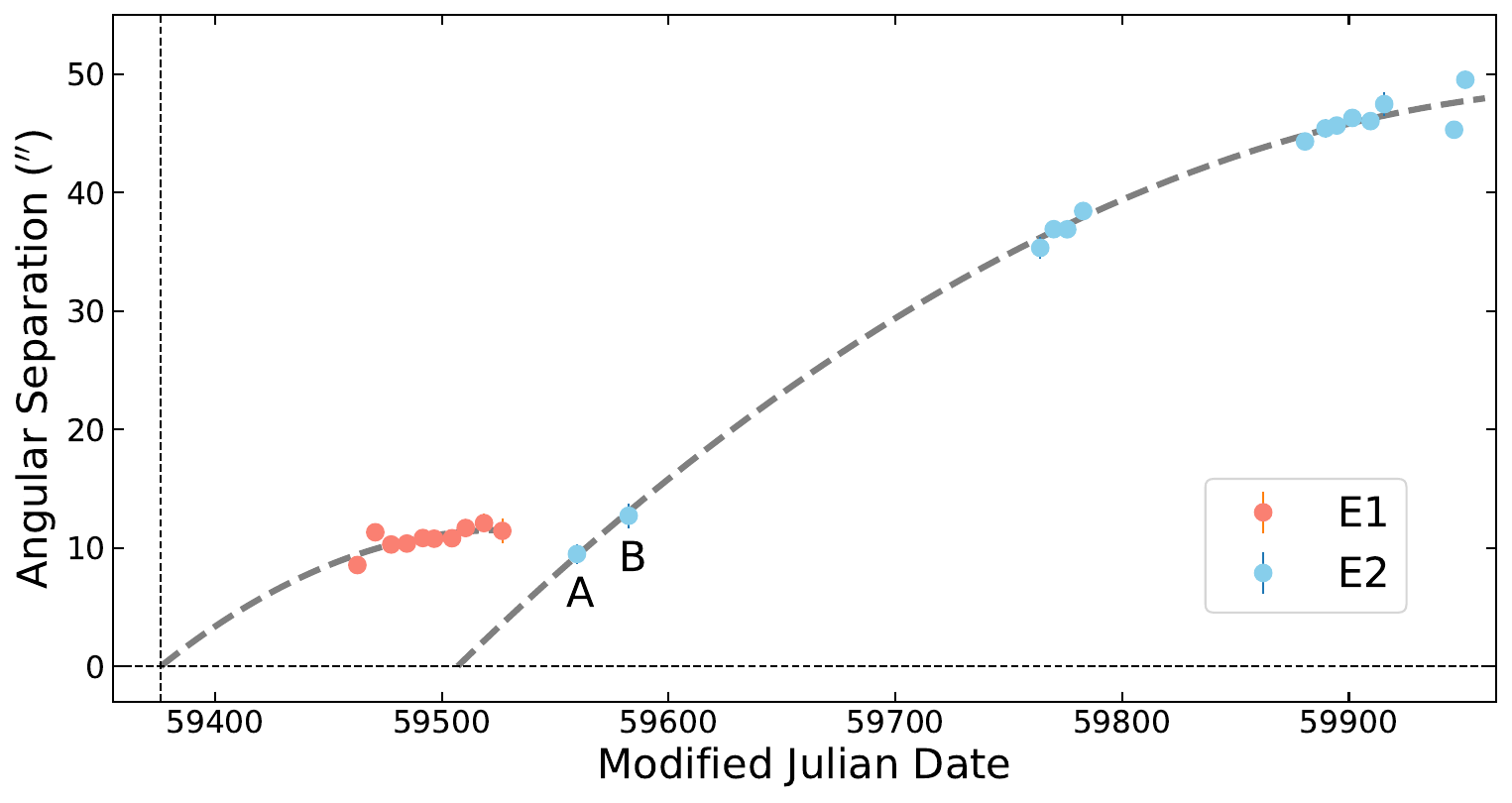}
 \caption{
 \footnotesize \textbf{ The classification of E1 (red) and E2 (blue) and two critical blobs  - Blob A and Blob B. }  Blob A marks the first discrete blob of E2, showing a steeper slope with Blob B and later blobs than those blobs detected at earlier times, indicating the occurrence of E2 following E1. }
 \label{fig:classification}
\end{figure*}
\clearpage

\newpage

\begin{figure*}
 \centering
 \includegraphics[width=0.8\textwidth]{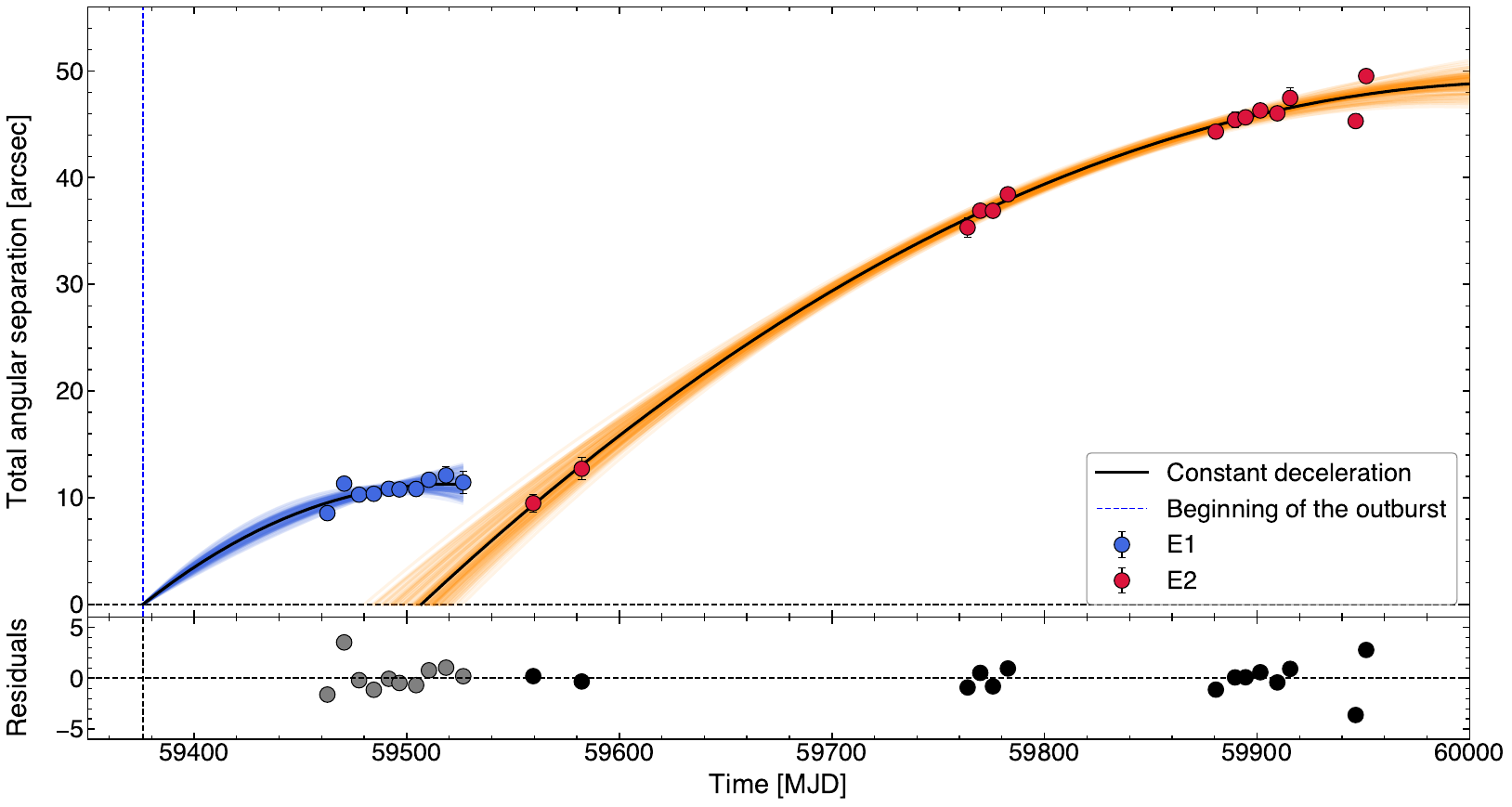}
 \caption{
 \footnotesize \textbf{Angular separation between the ejecta and the position of 4U~1543--47 with MCMC modelings.} Red and blue points are shown for E1 and E2, respectively, and the black continuous lines represent the best fits obtained with the constant deceleration models, with a fixed ejection date for E1, and a free ejection date for E2. The orange and blue shaded areas represent the total uncertainty in both fits and it is obtained by plotting the jet trajectories corresponding to the final positions of the MCMC walkers in the model parameter space. Residuals ([data – model]/uncertainties) are reported in the bottom panel. }
 \label{fig:angsep_mcmc}
\end{figure*} 
\clearpage

\newpage

\begin{figure*}
 \centering
 \includegraphics[width=0.40\textwidth]{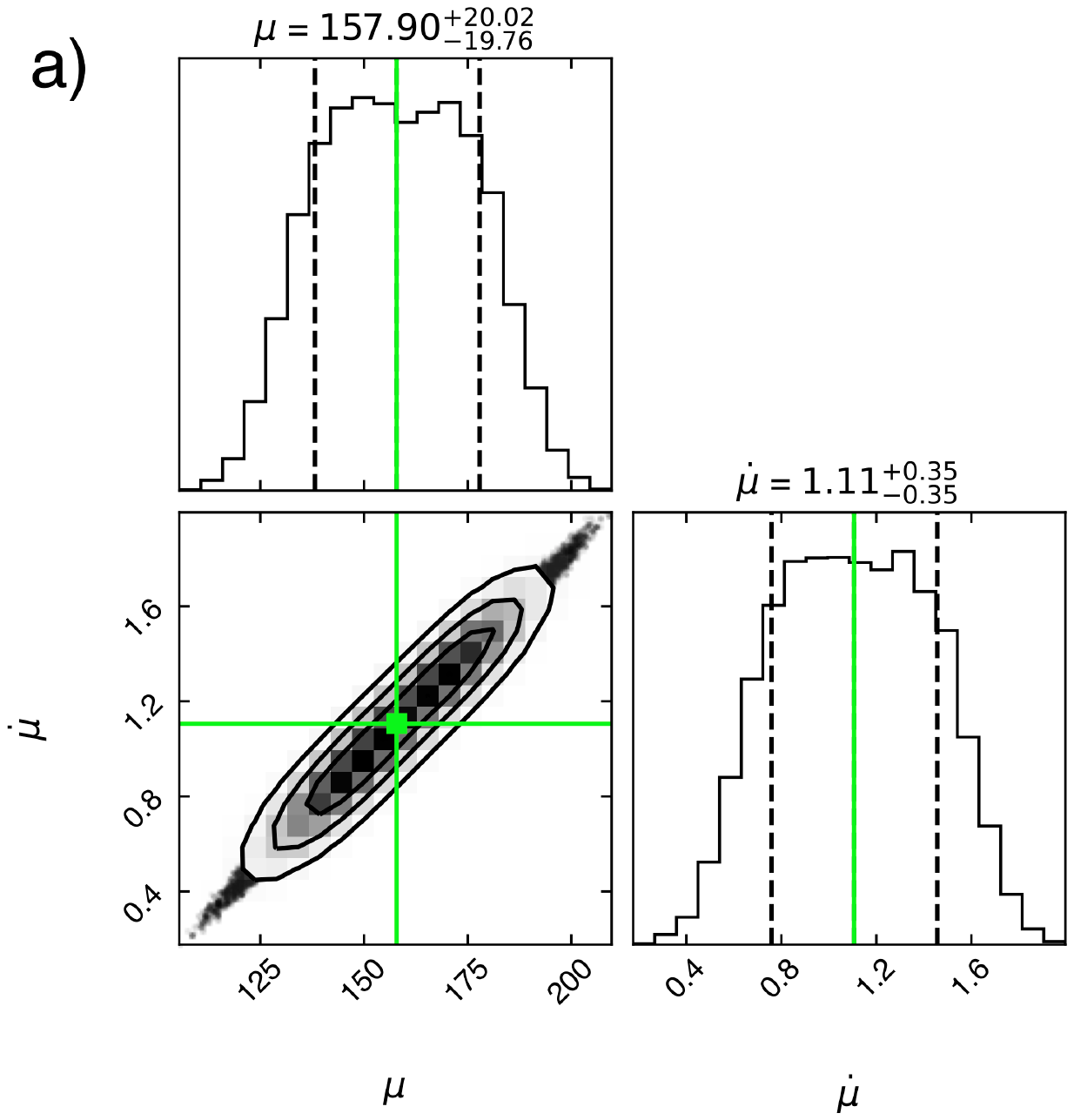}
 \includegraphics[width=0.40\textwidth]{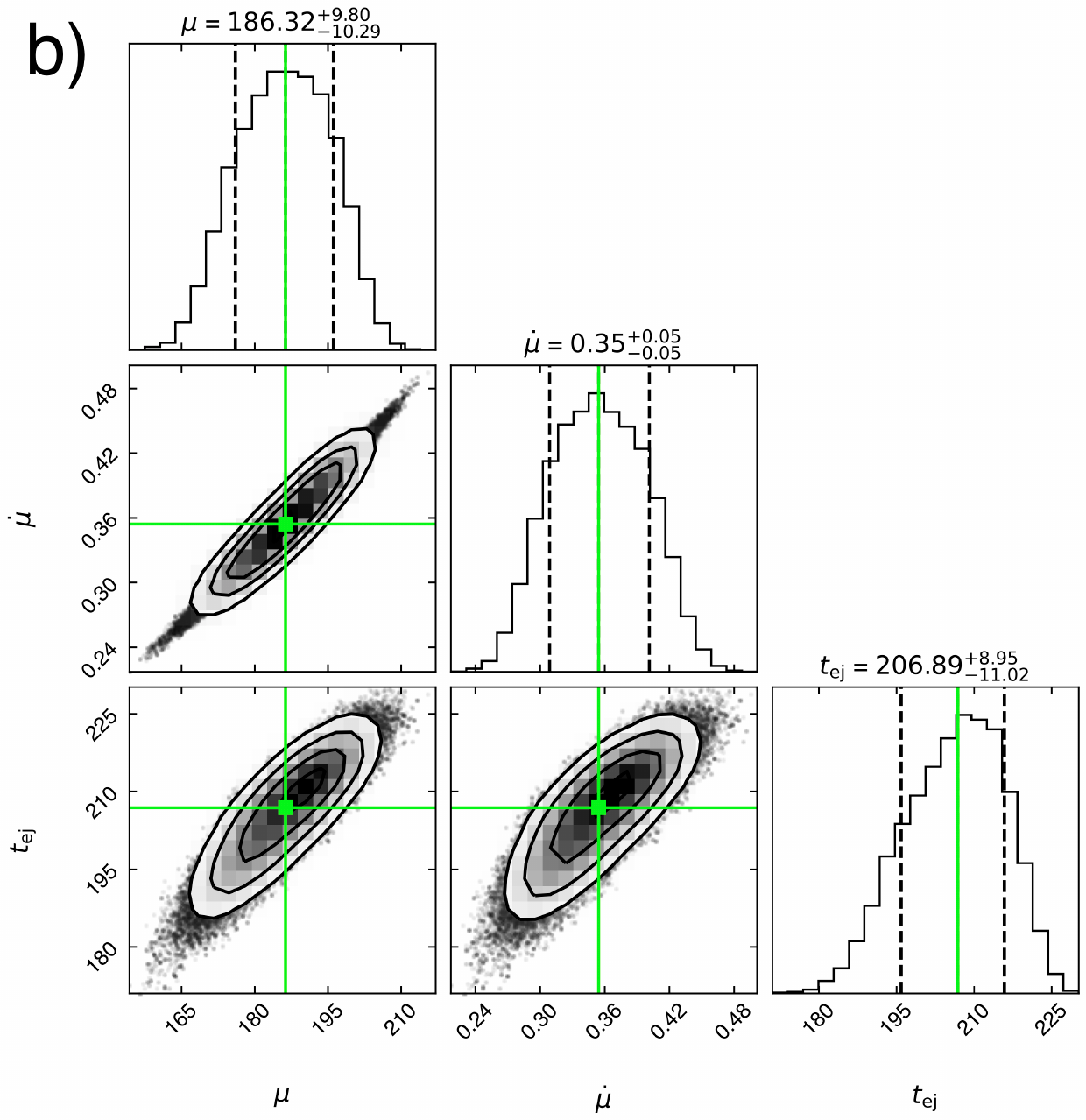}
 \caption{\footnotesize \textbf{Corner plots showing the constraints on the decelerated motion for the ejections.} The panels on the diagonal show histograms of the one dimensional posterior distributions for the model parameters, including the jet proper motion $\mu$ in mas/day, its time derivative $\dot{\mu}$ in mas/day$^2$ and the ejection date $t_{\rm ej}$, reported as $\sim$~MJD 59,506. The median value and the equivalent 1$\sigma$ uncertainty are marked with vertical dashed black lines. The other panels show the 2-parameter correlations, with the best-fit values of the model parameters indicated by green lines/squares. a) Modeled parameters of E1 (with a fixed ejection date). b) Modeled parameters of E2 (with an ejection date free to vary). }
 \label{fig:mcmc-corner}
\end{figure*}
\clearpage

\newpage

\begin{figure*}
 \centering
 \includegraphics[width=0.75\textwidth,angle=0]{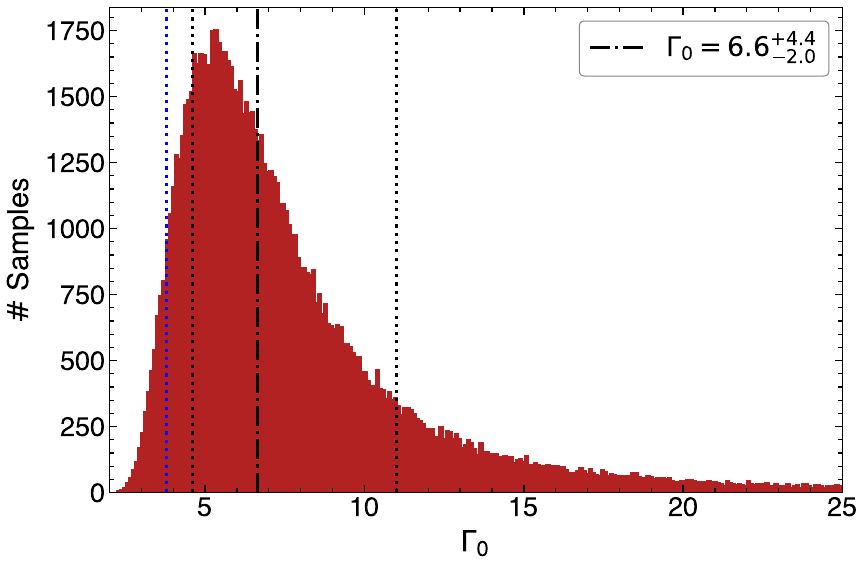}
 \caption{
 \footnotesize \textbf{Posterior distributions for the Lorentz factors derived for E1.} This is obtained by randomly sampling from the distributions of distance, proper motion and inclination angle (see text), which is based on a simple constant deceleration model for E1 with the launching time fixed. The legend shows the median value of $\Gamma_0$ (black dot-dashed vertical line) along with the 1$\sigma$ uncertainties (vertical black dotted lines), reported as the difference between the median and the 16th percentile of the posterior (lower error bar), and the difference between the 84th percentile and the median (upper error bar). The vertical blue dotted line marks the value of $\Gamma_0$ above which we find the 95\% of the samples. }
 \label{fig:mcmc-gamma}
\end{figure*}
\clearpage

\newpage
\begin{table}
\centering
\resizebox{0.9\textwidth}{!}{\begin{tabular}{ccccccccc}
\hline

{\bf Date} & {\bf MJD} & {\bf Flux density} & {\bf R.A.} & {\bf Dec.} & \multicolumn{2}{c}{\bf Error} & {\bf Beam Position Angle} & {\bf Angular Distance}\\
& (J2000) & ($\mu$Jy) & & & R.A. ($^{\prime\prime}$) & Dec. ($^{\prime\prime}$) & ($^{\circ}$) & ($^{\prime\prime}$)\\
\hline
2021 Sep 05 & 59,462.6971 & 111$\pm$18 & 15h47m08.777s & -47d40m17.19s & 0.56 & 0.70 & 5.09 & 8.55$\pm$0.61 \\
\hline
2021 Sep 13 & 59,470.5948 & 233$\pm$25 & 15h47m08.910s & -47d40m19.62s & 0.35 & 0.43 & -24.73 &11.32$\pm$0.38 \\
\hline
2021 Sep 20 & 59,477.6122 & 341$\pm$22 & 15h47m08.918s & -47d40m18.28s & 0.22 & 0.28 & -14.02 & 10.29$\pm$0.24 \\
\hline
2021 Sep 27 & 59,484.5291 & 365$\pm$23 & 15h47m08.916s & -47d40m18.39s & 0.23 & 0.26 & -35.25 & 10.37$\pm$0.22 \\
\hline
2021 Oct 04 & 59,491.5916 & 178$\pm$22 & 15h47m08.961s & -47d40m18.62s & 0.41 & 0.53 & -6.02 & 10.83$\pm$0.44 \\
\hline
2021 Oct 09 & 59,496.5915 & 170$\pm$22 & 15h47m08.950s & -47d40m18.64s & 0.43 & 0.56 & -3.65 & 10.78$\pm$0.47 \\
\hline
2021 Oct 17 & 59,504.4557 & 129$\pm$19 & 15h47m08.953s & -47d40m18.67s & 0.50 & 0.58 & -30.08 & 10.82$\pm$0.49 \\
\hline
2021 Oct 23 & 59,510.4664 & 108$\pm$19 & 15h47m09.045s & -47d40m19.01s & 0.57 & 0.71 & -19.36 & 11.68$\pm$0.59 \\
\hline
2021 Oct 31 & 59,518.5498 & 69$\pm$18 & 15h47m09.197s & -47d40m18.02s & 0.86 & 1.04 & 23.62 & 12.09$\pm$0.80 \\
\hline
2021 Nov 08 & 59,526.6119 & 77$\pm$23 & 15h47m09.129s & -47d40m17.81s & 1.53 & 1.09 & 73.77 & 11.44$\pm$1.06 \\
\hline
\hline
2021 Dec 11 & 59,559.5363 & 98$\pm$23 & 15h47m08.876s & -47d40m17.59s & 1.16 & 0.85 & 73.22 & 9.48$\pm$0.82 \\
\hline 
2022 Jan 03 & 59,582.3415 & 65$\pm$19 & 15h47m09.044s & -47d40m20.36s & 0.92 & 1.22 & 4.94 & 12.71$\pm$1.04 \\
\hline
2022 Jul 03 & 59,763.8626 & 74$\pm$20 & 15h47m10.555s & -47d40m37.09s & 0.89 & 1.12 & 13.38 & 35.33$\pm$0.93 \\
\hline
2022 Jul 09 & 59,769.8164 & 182$\pm$18 & 15h47m10.595s & -47d40m38.81s & 0.28 & 0.38 & 1.00 & 36.91$\pm$0.32 \\
\hline
2022 Jul 15 & 59,775.7679 & 148$\pm$23 & 15h47m10.654s & -47d40m38.30s & 0.44 & 0.60 & -10.84 & 36.90$\pm$0.49 \\
\hline
2022 Jul 22 & 59,782.8094 & 137$\pm$23 & 15h47m10.691s & -47d40m40.00s & 0.50 & 0.65 & 13.16 & 38.44$\pm$0.55 \\
\hline
2022 Oct 28 & 59,880.6445 & 167$\pm$26 & 15h47m11.030s & -47d40m44.79s & 0.71 & 0.53 & 73.91 & 44.31$\pm$0.51 \\
\hline
2022 Nov 06 & 59,889.6495 & 111$\pm$25 & 15h47m11.196s & -47d40m44.83s & 1.16 & 0.76 & 80.38 & 45.42$\pm$0.77 \\
\hline
2022 Nov 11 & 59,894.6079 & 133$\pm$22 & 15h47m11.243s & -47d40m44.74s & 0.72 & 0.57 & 67.38 & 45.66$\pm$0.54 \\
\hline
2022 Nov 18 & 59,901.5453 & 122$\pm$24 & 15h47m11.215s & -47d40m45.82s & 0.69 & 0.71 & 42.52 & 46.30$\pm$0.62 \\
\hline
2022 Nov 26 & 59,909.5446 & 124$\pm$25 & 15h47m11.390s & -47d40m43.90s & 0.79 & 0.72 & 53.8 & 46.03$\pm$0.64 \\
\hline
2022 Dec 02 & 59,915.5835 & 138$\pm$40 & 15h47m11.297s & -47d40m46.65s & 1.54 & 0.96 & 82.78 & 47.47$\pm$1.00 \\
\hline
2023 Jan 02 & 59,946.4028 & 123$\pm$26 & 15h47m11.234s & -47d40m44.34s & 0.68 & 0.77 & 30.85 & 45.30$\pm$0.65 \\
\hline
2023 Jan 07 & 59,951.3464 & 107$\pm$21 & 15h47m11.481s & -47d40m47.77s & 0.57 & 0.74 & 10.61 & 49.52$\pm$0.62 \\
\hline
\end{tabular}}
\caption{
\textbf{Angular separations of all discrete blobs measured with MeerKAT Observations.} The measurements of blobs for E1 and E2 with MeerKAT observations are arranged in time sequence and are separated by double horizontal lines. It is worth noting that the first detection of E2 has a smaller angular distance from the core by 2.4 $\sigma$ than the average distance of the last three detections of E1, independently supporting a subsequent second ejection occurred.  }
\label{tab:meerkat_observations}
\end{table}

\clearpage

\newpage

\section*{Supplementary Information}
\renewcommand{\baselinestretch}{1.0}
\selectfont

\subsection{Contour Images of the Detected 24 Discrete Blobs}
In \SUPFIG{fig:contour_img}, we show the MeerKAT images of all the detected 24 discrete blobs at 1.28 GHz. In these images, the Gaia DR2 optical position (i.e., core position) of \target~is marked by a black cross and the radio positions of the discrete blobs, to the South-East direction of the core position, are marked by the red crosses. The contours in all the images are chosen to correspond to (-1, 1, $\sqrt{2}$, 2, 3, 4, 5, 6, 7, 8, 9, 10, 20, 30, 40, 50, 80, 100, 200, 500, 800, 1000) times the 3$\times$rms noise, respectively, for the purpose of the presentation of individual detection and visualization of the complete motions of the blobs in the image plane.

\noindent
\subsection{Classifications of Discrete Blobs and Alternative Model Fits}
Classifications of discrete blobs and the corresponding model fits favored by the data are presented in the Main text and the Method section. Here we present details of alternative model fits, which help clarify the most likely classification of the ejection blobs and critical measurements of Blob A and Blob B. 

We have applied three rules (see Methods) in understanding the proper motion of the blobs, namely, the ``Going Forward" rule, the ``No Acceleration" rule, and the ``Least Ejection" rule, when classified the 24 discrete blobs into two ejections. Investigation of an individual blob or a group of blobs with the ``Going Forward" rule and the ``No Acceleration" rule have helped us determine that there were more than one ejection that launched from our source. Blob A is most likely the first appearance of the propagating blob corresponding to the second ejection E2, by showing a smaller angular separation from the core, higher proper motion than the previous blobs, and occurred at a later time. Alternative classifications of the ejection blobs as E1 or E2 are theoretically possible, but less probable, as suggested by the details of model fits of various considerations introduced below. 

Following the ``Least Ejection" rule, we started to model the 24 blobs as due to a single ejection.  This means that those blobs are actually the same blob but detected at different times and at different positions (i.e., same blob in different images) when it was moving forward. We applied either a constant deceleration model or a ballistic motion model to trace the motion of the blobs, which results in a reduced $\chi^2$ of 13.5 (dof$=$22). Therefore a single ejection scenario is not able to describe the motion of the blobs. The most significant residuals between the data and the models come from the deceleration trend of E1 and the deviations of Blob A and B from the trend established by the previous E1 blobs, by showing an increase in the speed of the proper motion. These deviations also violate the "Moving Forward" rule and the "No Acceleration" rule. In conclusion, a second ejection E2 in addition to the first ejection E1 is required to model the proper motion of the blobs. An example of the model fits with a single ejection is shown as \SUPFIG{fig:ang_sep_linearfit}. 

Next, we consider to model the proper motions of the blobs with a two-ejection scenario in which E1 and E2 correspond to the first and the second ejection, respectively. Here below we show alternative classifications of the blobs under the two-ejection scenario of which either Blob A or Blob B, or both, do not belong to E2. Here we summarize our results. 

\begin{itemize}
    \item  Case 1: Blob A is in E1 and Blob B is in E2 (\SUPFIG{fig:ang_sep_AE1_BE2}): In this case, we obtain an initial proper motion of $\mu=$136.6$\pm$5.9 mas/day with a reduced $\chi^{2} \sim$~4.3 (dof$=$9) for the newly classified first ejection, with the launch time fixed to MJD 59,376. We also obtain an initial proper motion of $\mu=$188.6$\pm$7.9 mas/day with a reduced $\chi^{2} \sim$~2.6 (dof$=$10) for the newly classified second ejection, while the fitted launch time is MJD 59,509.3$\pm$9.0. The residual corresponding to Blob A is -3.7 $\sigma$, corresponding to 3.1\arcsec.
    
    \item Case 2: Blob A is in E2 and Blob B is in E1 (\SUPFIG{fig:ang_sep_AE2_BE1}): In this case, we obtain an initial proper motion of $\mu=$131.3$\pm$5.4 mas/day with a reduced $\chi^{2} \sim$~3.2 (dof$=$9) for the newly classified first ejection, with the launch time fixed to MJD 59,376. We also obtain an initial proper motion of $\mu=$185.6$\pm$5.8 mas/day with a reduced $\chi^{2} \sim$~2.6 (dof$=$10) for the newly classified second ejection, while the fitted launch time is MJD 59,505.5$\pm$6.3. The residual corresponding to the Blob B is -0.8 $\sigma$, corresponding to -0.8\arcsec.
    
    \item Case 3: Blob A and Blob B are both in E1 (\SUPFIG{fig:ang_sep_ABinE1}): In this case, we obtain an initial proper motion of $\mu=$129.7$\pm$4.7 mas/day with a reduced $\chi^{2} \sim$~5.1 (dof$=$10) for the newly classified first ejection, with the launch time fixed to MJD 59,376. We also obtain an initial proper motion of $\mu=$179.2$\pm$29.3 mas/day with a reduced $\chi^{2} \sim$~2.9 (dof$=$9) for the second ejection E2, while the fitted launch time is MJD 59,496.8$\pm$40.0. The residual for Blob A is -4.5 $\sigma$, corresponding to 3.7\arcsec, while the residual for Blob B is -0.6 $\sigma$, corresponding to -0.7\arcsec.
 
\end{itemize}

In the above cases, which associate Blob A and/or Blob B to E1 instead of E2, the model fits regarding E1 blobs have significantly larger reduced $\chi^{2}$ of $\sim$~4.3, $\sim$~3.2, and $\sim$~5.1, respectively, as compared to the reduced $\chi^{2}$ of $\sim$~2.4 for the blob association of E1 presented in the main text and Methods (i.e., both Blob A and Blob B are associated with E2); the model fits regarding E2 blobs without Blob A and/or Blob B have slightly larger reduced $\chi^{2}$ of $\sim$~2.6, $\sim$~2.6, and $\sim$~2.9, respectively, as compared to the reduced $\chi^{2}$ of $\sim$~2.4 for E2 presented in the main text and Methods. The possible association of Blob A with E1 is unlikely based on the large residuals in these alternative model fits to the E1 blobs. The possible association of Blob B with the E1 blobs can not be rejected, but unlikely, as it would require a non-detection of the moving Blob A (the E2 blob) when Blob B was detected. Furthermore, it is also worth noting that other possible classifications of the 24 blobs will require more than two ejections and not meaningful.

\clearpage

\bigskip\noindent
\SUPFIG{fig:contour_img}: {\bf Full contour images of the 24 discrete blobs as seen with MeerKAT.} The core position (black) and the discrete blobs (red) are marked with a cross. Contours are (-1, 1, $\sqrt{2}$, 2, 3, 4, 5, 6, 7, 8, 9, 10, 20, 30, 40, 50, 80, 100, 200, 500, 800, 1000) times the 3$\times$rms noise. 

\bigskip\noindent \SUPFIG{fig:ang_sep_linearfit}: {\bf A ballistic model fit to the angular separations of the 24 discrete blobs as a single ejection.} Those blobs corresponding to E1 and E2, respectively, are shown as red and blue circles. A ballistic model (dashed line) can not describe the motion of the blobs as a single ejection, yielding a reduced $\chi^2$ of 13.5 (dof$=$22). Residuals ([data – model]/error in data) are shown.

\bigskip\noindent 
\SUPFIG{fig:ang_sep_AE1_BE2}: {\bf Constant deceleration model fits to the E1 blobs and the E2 blobs, respectively, but with Blob A included in those E1 blobs.} Those blobs corresponding to E1 and E2 are shown as red and blue filled circles, respectively. In the two model fits, Blob A is included in the E1 blobs and Blob B is still in the E2 blobs. The residuals ([data – model]/error in data) are shown. The residual corresponding to Blob A is -3.7 $\sigma$, corresponding to 3.1\arcsec, which suggests it's unlikely belonging to E1.

\bigskip\noindent 
\SUPFIG{fig:ang_sep_AE2_BE1}: {\bf Constant deceleration model fits to the E1 blobs and the E2 blobs, with Blob B included in E1 and Blob A still included in E2.} Those blobs corresponding to E1 and E2 are shown as red and blue filled circles, respectively. The residuals ([data – model]/error in data) are shown. The residual corresponding to the Blob B is -0.8 $\sigma$, corresponding to -0.8\arcsec. If Blob B belongs to E1 and Blob A belongs to E2, then Blob A would not only being not detected when Blob B was detected but also Blob A should have passed Blob B eventually.

\bigskip\noindent 
\SUPFIG{fig:ang_sep_ABinE1}: {\bf Constant deceleration model fits to the E1 and the E2 blobs with both Blob A and Blob B included in the E1 blobs.} E1 and E2 are shown as red and blue filled circles. Residuals ([data – model]/error in data) are shown. The residual for Blob A is -4.5 $\sigma$, corresponding to 3.7\arcsec, while the residual for Blob B is -0.6 $\sigma$, corresponding to -0.7\arcsec, which suggests that Blob A is unlikely belonging to E1.

\clearpage

\renewcommand{\figurename}{\text{Supplementary $|$ Figure}}

\begin{figure*}
\centering
\includegraphics[width=0.90\textwidth,angle=0]{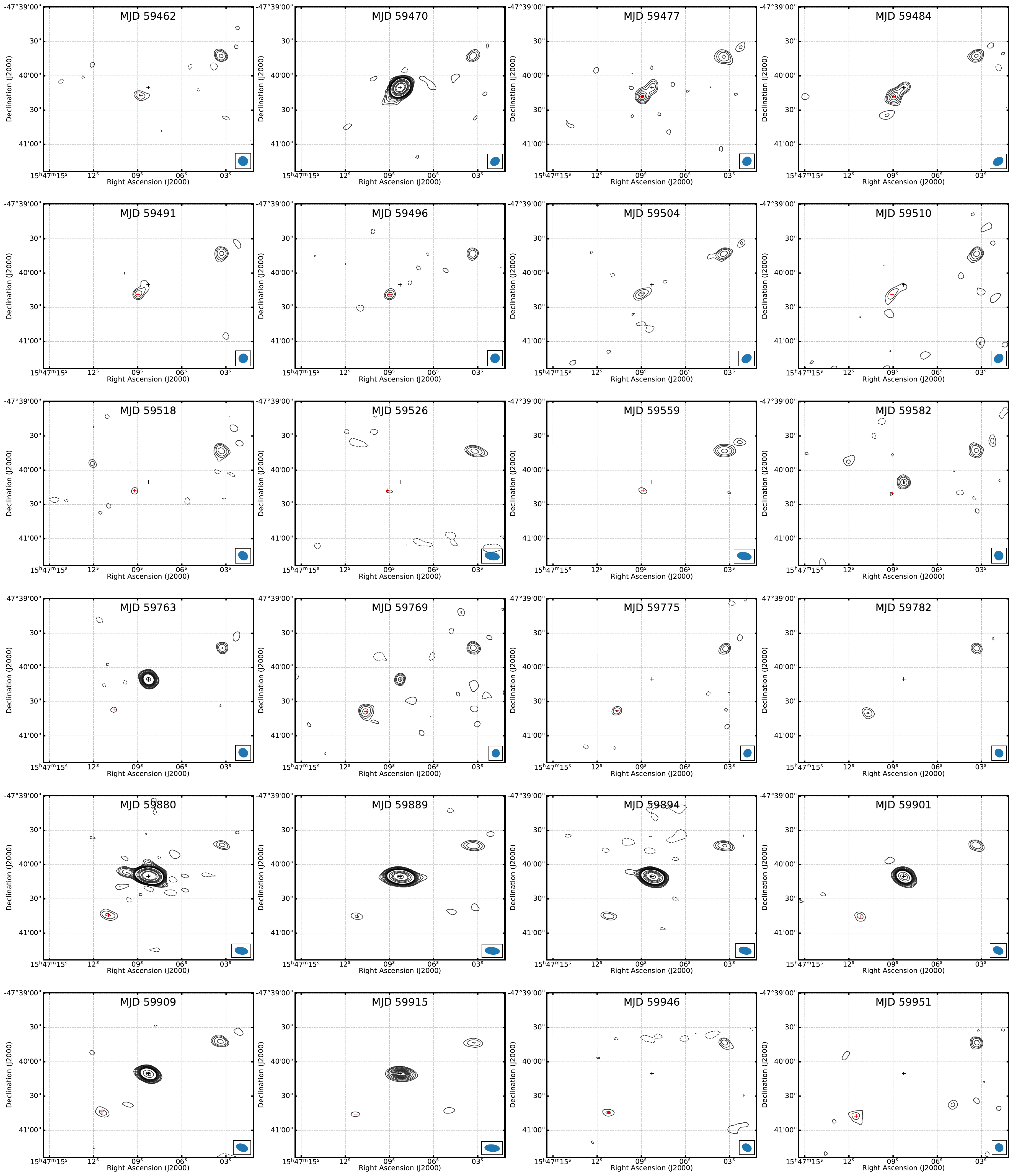}

\caption{
\footnotesize \textbf{Full contour images of the 24 discrete blobs as seen with MeerKAT.} The core position (black) and the discrete blobs (red) are marked with a cross. Contours correspond to -1 (in dashed line), 1, $\sqrt{2}$, 2, 3, 4, 5, 6, 7, 8, 9, 10, 20, 30, 40, 50, 80, 100, 200, 500, 800, and 1000 times the 3$\times$rms noise, respectively.}
\label{fig:contour_img}
\end{figure*}
\clearpage

\begin{figure*}
\centering
\includegraphics[width=0.90\textwidth,angle=0]{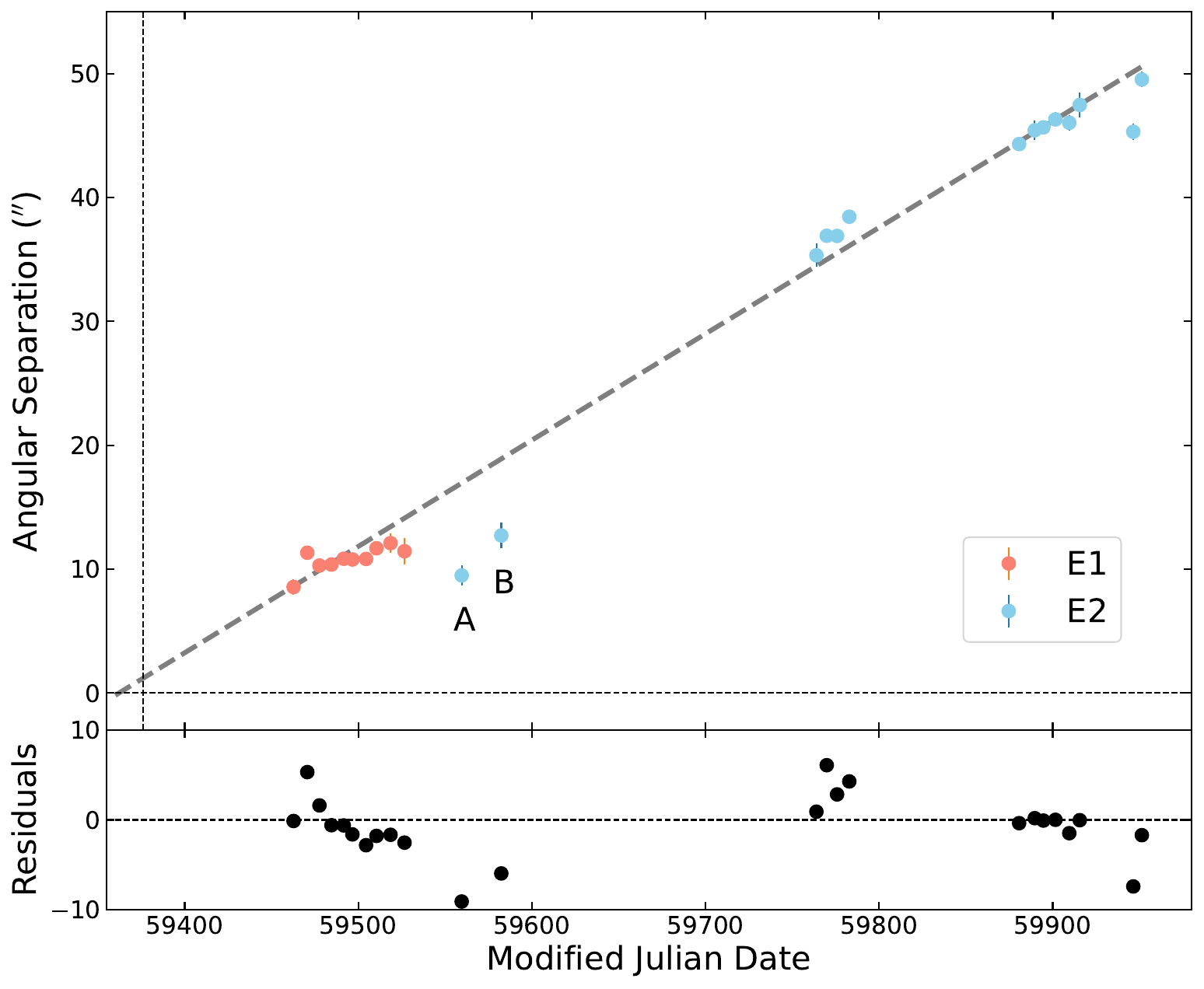}

\caption{
\footnotesize \textbf{A ballistic model fit to the angular separations of the 24 discrete blobs as a single ejection.} Those blobs corresponding to E1 and E2, respectively, are shown as red and blue circles. A ballistic model (dashed line) can not describe the motion of the blobs as a single ejection, yielding a reduced $\chi^2$ of 13.5 (dof$=$22). Residuals ([data – model]/error in data) are shown.}
\label{fig:ang_sep_linearfit}
\end{figure*}
\clearpage

\begin{figure*}
\centering
\includegraphics[width=0.90\textwidth,angle=0]{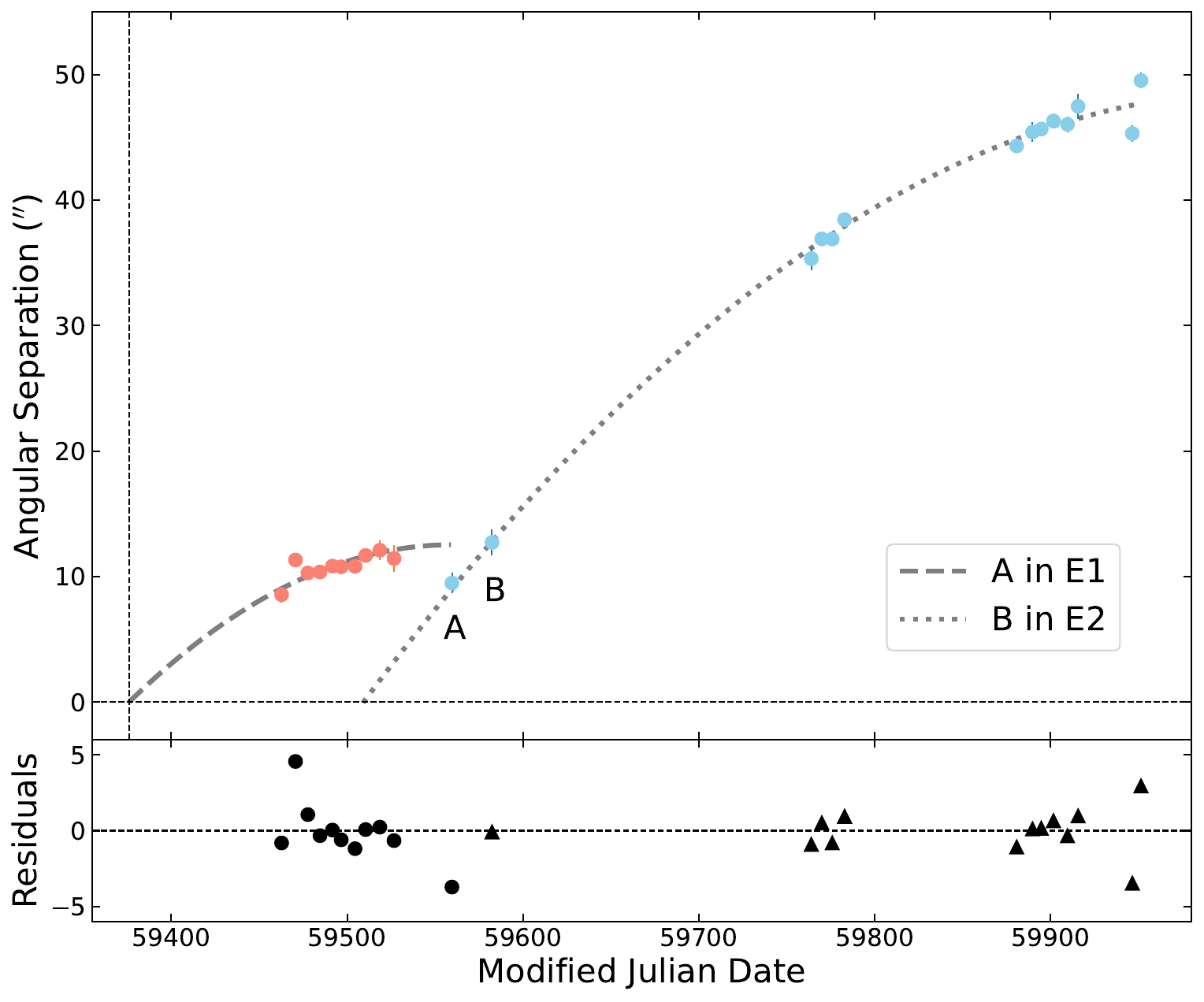}

\caption{
\footnotesize \textbf{Constant deceleration model fits to the E1 blobs and the E2 blobs, respectively, but with Blob A included in those E1 blobs.} Those blobs corresponding to E1 and E2 are shown as red and blue filled circles, respectively. In the two fits, Blob A is included in the E1 blobs and Blob B is still in the E2 blobs. The residuals ([data – model]/error in data) are shown. The residual corresponding to Blob A is -3.7, corresponding to 3.1\arcsec, which suggests it's unlikely belonging to E1. 
}
\label{fig:ang_sep_AE1_BE2}
\end{figure*}
\clearpage

\begin{figure*}
\centering
\includegraphics[width=0.90\textwidth,angle=0]{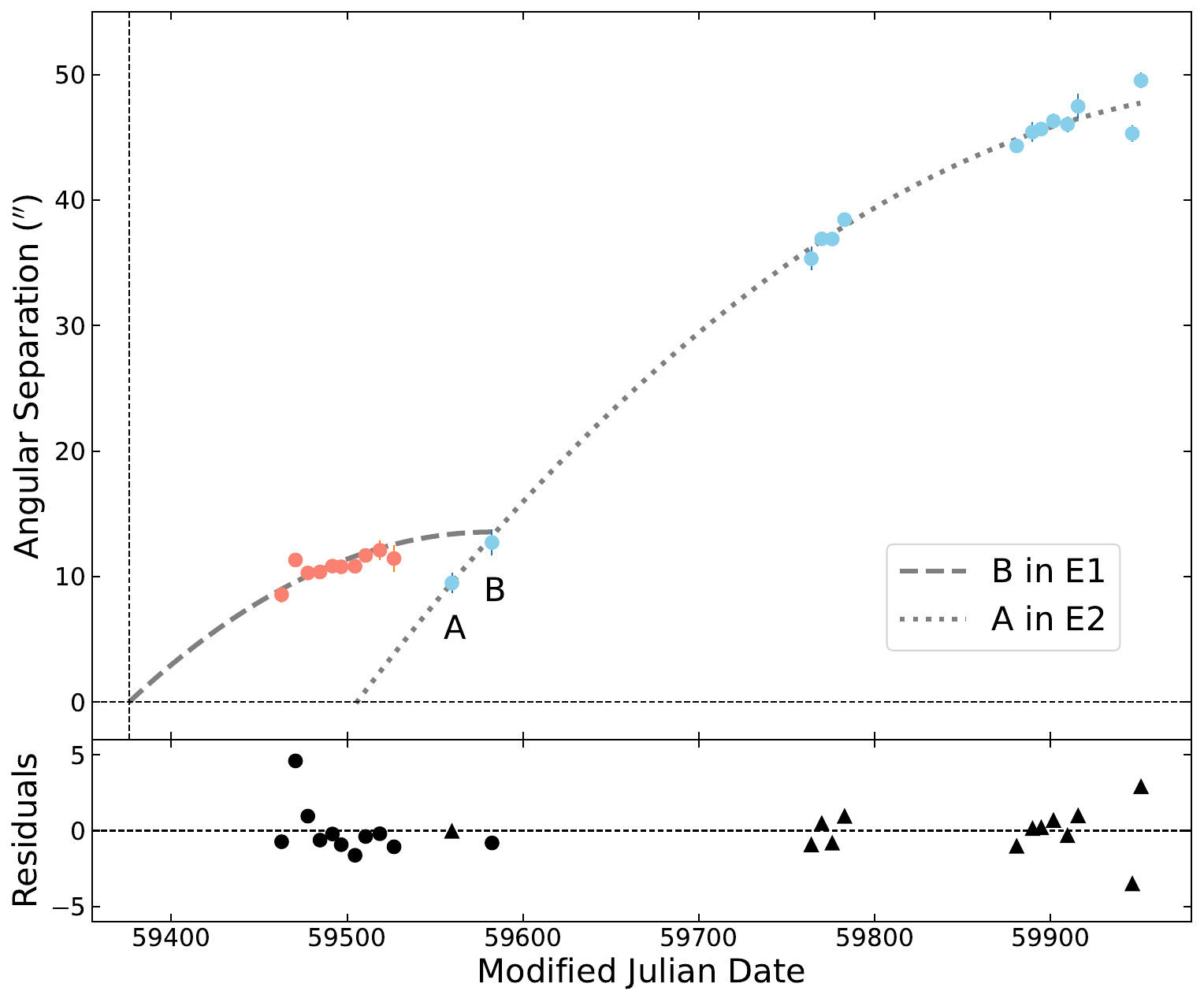}

\caption{
\footnotesize \textbf{Constant deceleration model fits to the E1 blobs and the E2 blobs, with Blob B included in E1 and Blob A still included in E2.} Those blobs corresponding to E1 and E2 are shown as red and blue filled circles, respectively. The residuals ([data – model]/error in data) are shown. The residual corresponding to the Blob B is -0.8, corresponding to -0.8\arcsec. If Blob B belongs to E1 and Blob A belongs to E2, then Blob A would not only being not detected when B was detected but also Blob A should have passed the Blob B eventually.}
\label{fig:ang_sep_AE2_BE1}
\end{figure*}
\clearpage

\begin{figure*}
\centering
\includegraphics[width=0.90\textwidth,angle=0]{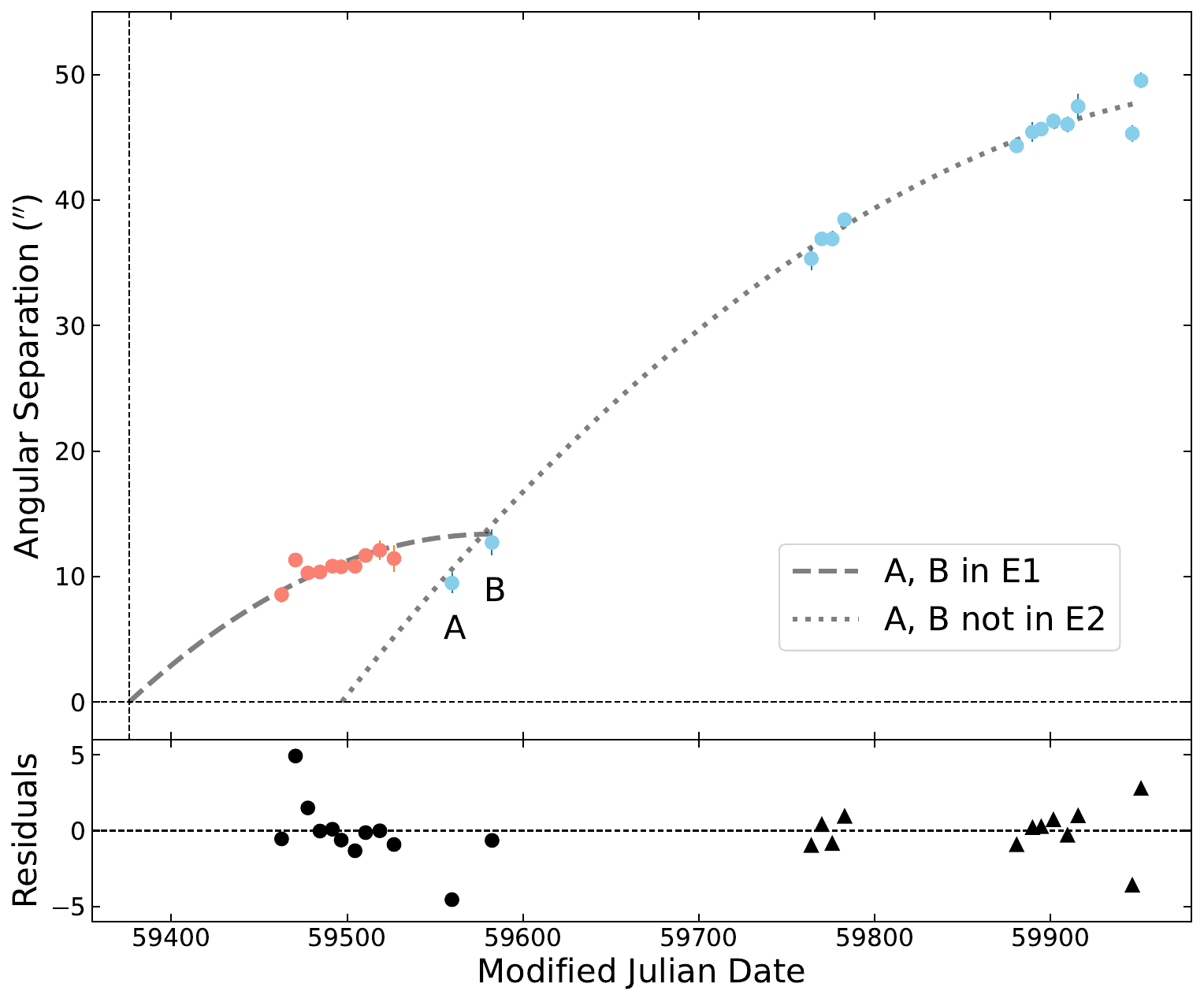}

\caption{
\footnotesize \textbf{Constant deceleration model fits to the E1 and the E2 blobs with both Blob A and Blob B included in the E1 blobs.} E1 and E2 are shown as red and blue filled circles. Residuals ([data – model]/error in data) are shown. The residual for Blob A is -4.5, corresponding to 3.7\arcsec, while the residual for Blob B is -0.6, corresponding to -0.7\arcsec, which suggests that Blob A is unlikely belonging to E1. }
\label{fig:ang_sep_ABinE1}
\end{figure*}
\clearpage

\end{document}